\documentclass[9.5pt, journal, compsoc]{IEEEtran}
\usepackage[switch]{lineno}
\usepackage{amsmath,amssymb,amsfonts,amsthm}
\usepackage{breqn}

\usepackage{graphicx}
\usepackage{xcolor}
\usepackage{xspace}
\usepackage{hyperref}
\usepackage{booktabs}
\usepackage[linesnumbered,ruled,vlined]{algorithm2e}
\SetAlFnt{\small}
\SetAlCapFnt{\large}
\SetAlCapNameFnt{\large}

\usepackage{multirow}

\usepackage{balance}
\usepackage{comment}
\usepackage{ifthen}
\usepackage{subcaption}
\usepackage[utf8]{inputenc}
\usepackage[T1]{fontenc}

\newcommand{\crawljax}{\textsc{Crawljax}\xspace}  
\newcommand*\rot{\rotatebox{90}}

\SetCommentSty{mycommfont}

\emergencystretch=\maxdimen

\usepackage{listings}
\usepackage{xcolor}
 
\definecolor{codegreen}{rgb}{0,0.6,0}
\definecolor{codegray}{rgb}{0.5,0.5,0.5}
\definecolor{codepurple}{rgb}{0.58,0,0.82}
\definecolor{backcolour}{rgb}{0.95,0.95,0.92}

\newboolean{showcomments}
\setboolean{showcomments}{true}
\ifthenelse{\boolean{showcomments}}
{
	\definecolor{myyellow}{RGB}{255, 228, 26}
	\definecolor{myblue}{RGB}{50, 50, 220}
	\newcommand{\nb}[2]{
		{\sf
			\fcolorbox{myyellow}{yellow}{\scriptsize\textbf{#1}}%
			$\blacktriangleright$%
			{\color{orange}\fontsize{8pt}{8pt}\selectfont\textbf{#2}}%
		}%
	}
}
{
	\newcommand{\nb}[2]{}
}

\newcommand{\ndOne}{Nd$_1$-\emph{data}\xspace}
\newcommand{\ndTwo}{Nd$_2$-\emph{data}\xspace}
\newcommand{\ndThree}{Nd$_3$-\emph{struct}\xspace}

\newcommand{\ali}[1]{\nb{Ali}{#1}}

\newcommand{\model}{$\mathcal{M}$}
\newcommand{\StateVertex}[2][]{$\mathcal{S}_{#2}^{#1}$}
\newcommand{\Fragment}[2][]{$\mathcal{F}_{#2}^{#1}$}

\newcommand{\header}[1]{\par\smallskip\noindent\textbf{#1.}}

\newcommand{\transition}[1]{$\mathcal{A}_{#1}$}
\newcommand{\Path}[2][]{$\mathcal{P}_{#2}^{#1}$}
\newcommand{\actionable}[1]{$\alpha_{#1}$}

\newcommand{\testCase}[1]{$\mathcal{T}_{#1}$}
\theoremstyle{definition}
\newtheorem{definition}{Definition}
\newcommand{\q}[1]{RQ\textsubscript{#1}}

\newcommand{\toolname}{\textsc{FragGen}\xspace}

\usepackage{color}
\usepackage{textcomp}
\definecolor{listinggray}{gray}{0.9}
\definecolor{lbcolor}{rgb}{0.9,0.9,0.9}
\makeatletter
\DeclareRobustCommand{\change}{%
  \@bsphack
  \leavevmode
  \normalcolor
  \@esphack
}
\DeclareRobustCommand{\stopchange}{%
  \@bsphack
  \normalcolor
  \@esphack
}
\makeatother

\usepackage[inline]{enumitem}

\def\BibTeX{{\rm B\kern-.05em{\sc i\kern-.025em b}\kern-.08em
    T\kern-.1667em\lower.7ex\hbox{E}\kern-.125emX}}

\begin{document}
\thispagestyle{plain}
\pagestyle{plain}
\title{
Fragment-Based Test Generation For Web Apps
}


\author{Rahulkrishna Yandrapally, \IEEEmembership{Student Member, IEEE,} and
                Ali Mesbah,~\IEEEmembership{Member, IEEE} \\
}

\maketitle
\begin{abstract}

Automated model-based test generation presents a viable alternative to the costly manual test creation currently employed for regression testing of web apps. 
However, existing model inference techniques rely on threshold-based whole-page comparison to establish state equivalence, which cannot 
reliably identify near-duplicate web pages in modern web apps. Consequently, existing techniques produce inadequate models for dynamic web apps, and fragile test oracles, 
 rendering the generated regression test suites ineffective. 
We propose a model-based test generation technique, \toolname, that eliminates the need for thresholds, by employing a novel state abstraction based on page fragmentation to establish state equivalence. 
\toolname also uses fine-grained page fragment analysis to diversify state exploration and generate reliable test oracles.
Our evaluation shows that \toolname outperforms existing whole-page techniques by detecting more near-duplicates, inferring better web app models and generating test suites that are better suited for regression testing. 
\change
 On a dataset of 86,165 state-pairs, \toolname detected 123\% more near-duplicates on average compared to whole-page techniques.\stopchange{} The crawl models inferred by \toolname have 62\% more precision and 70\% more recall on average.
  \toolname also generates reliable regression test suites with test actions that have nearly 100\% success rate on the same version of the web app even if the execution environment is varied.
 The test oracles generated by \toolname can detect 98.7\% of the visible changes in web pages while being highly robust, making them suitable for regression testing.
\end{abstract}

 \section{introduction}
 Regression testing of modern web apps is a costly activity~\cite{6606562} in practice, which requires developers to manually create  test suites, using a combination of programming and record/replay tools such as Selenium~\cite{roest_regression_2010}. 
   In addition, maintaining such test suites is known to be costly \cite{memon11regression, Grechanik:2009:MEG:1555001.1555055}  as even minor changes of the app can cause many tests to break; for example, according to a study at Accenture~\cite{Grechanik:2009:MEG:1555001.1555055} even simple modifications to the user interface  of apps result in 30--70\% changes to tests, which costs \$120 million per year to repair. When the test maintenance cost becomes overwhelming, whole test suites are abandoned~\cite{Christophe2014}.
    

  Given the short release cycles of modern web apps and maintenance costs of manually written tests,  
   automatic generation of regression test suites seems a viable alternative. However, the effectiveness of web test generation techniques~\cite{2019-Biagiola-FSE-Diversity, mesbah:tse12, mirshokraie2012jsart} is limited by the ability to obtain an accurate and complete model of the app under test. 
   Manual construction of such models for complex apps is not practical. 
   Automated model inference techniques~\cite{mesbah:tweb12} trigger user actions such as clicking on buttons and record corresponding transitions between states in the web app to build a graph-based model. 

   One particular  challenge here is the presence of \emph{near-duplicate} states in web apps, which can adversely impact the  inferred model in terms of redundancy and adequacy~\cite{ndStudy2020}. 
  Near-duplicates are states that are \emph{similar} to each other in terms of functionality~\cite{Henzinger:2006:FNW:1148170.1148222}.  


  
Another important challenge is the generation of test assertions. Two factors directly contribute to the challenge here, namely effectiveness and tolerance/robustness, i.e., regression test assertions should be able to detect unexpected app behavior, but at the same time, be tolerant to minor changes that do not affect the functionality. 

Both model inference and test oracle generation thus require suitable abstractions to produce effective and robust test suites.
 Existing techniques generate regression tests that compare the \emph{whole} page as seen during testing with an instance of the page recorded on a previous version~\cite{roest_regression_2010, mesbah:tse12}. Our insight is that, such whole-page comparison techniques, although effective at detecting changes, are not tolerant enough to handle near-duplicates and make the test suites fragile. 
  Test fragility is known to be a huge problem in web testing~\cite{2015-leotta-ICST}. 

In this work, we propose a novel state abstraction that employs fine-grained fragments to establish functional equivalence.  We conjecture that a web page is not a singular functional entity and thus partitioning it into separate   fragments can help in determining functional equivalency when comparing different states.  Using this novel state abstraction, we have developed a technique, called \toolname. 
\change Our fragment-based analysis enables us to  (1) prioritize available actions to diversify exploration, 
(2) accomplish state comparison without the need for manually selecting thresholds, a manual tedious fine-tuning process, required for all existing techniques~\cite{ndStudy2020};  our state comparison algorithm leverages both \emph{structural} and \emph{visual} properties of the page fragments to 
identify near-duplicate characteristics
specific to the web app under test during model inference, and (3) generate  test assertions that operate at the fragment level instead of the whole page level,  and  apply fragment \emph{memoization} to make them much more robust to state changes 
  that should not break regression tests.
\stopchange{}

  \begin{figure*}[h]
\centering
\includegraphics[trim=0cm 0cm 0cm 0cm, clip=true, width=0.99\linewidth]{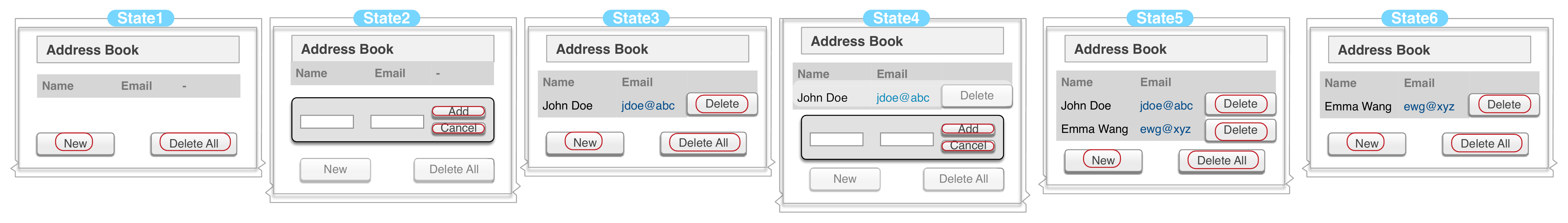}
\caption{Motivating example: app states with \emph{actionables} highlighted. 
}
\label{fig:fig_near-duplicates}
\vspace{-0.5em}
\end{figure*}
 
Our empirical evaluation shows that \toolname is able to outperform whole-page techniques in classifying state-pairs and identifying near-duplicates. \change On a dataset of 86,165 manually labelled state-pairs, \toolname detected 123\% more near-duplicates on average and 82\% more than the best performing existing technique.\stopchange{} When employed to infer web app models, \toolname is able to 
  diversify exploration using fine-grained fragment analysis to produce models with 70\% higher precision and 62\% higher recall on an average compared to the state-of-the-art technique.
 In addition, our evaluation shows that \toolname generated test suites that are better suited for regression testing. 
Where existing techniques generated brittle test suites with nearly 17\% test actions that fail even on the same version of the web app, 
 \toolname generated reliable test suites with nearly 100\% successful test actions on the same version of the web app, and detected more app changes with fewer false positives when run on different application versions. 
  When evaluated through mutation analysis, \toolname's test oracles 
  could detect 98.7\% of visible state changes while being tolerant enough to ignore 90.6\% of equivalent mutants.
 

 
This paper makes the following contributions:
\begin{itemize}
 \item A technique for determining state equivalence (i.e., distinct, clone, near-duplicate) at the fragments-level, without a need for setting thresholds.
 \item A state abstraction technique that uses the structural as well as visual properties of fragments for equivalence checking. 
 \item A novel model inference approach that employs page fragments to explore the
 web application state-space.
 \item A fine-grained fragment-based test assertion generation for effective and reliable regression testing. 
 \item The evaluation and implementation of  \toolname, which is publicly available~\cite{toolLink}. 
\end{itemize}

\section{Background and Motivation}
\change In this section, we provide the background information on web app testing and analysis, and introduce key terms and concepts that are used in the rest of the paper. \stopchange

We use a running example, shown in~\autoref{fig:fig_near-duplicates},   based on one of our subject systems~\cite{addressbook}. 
The example web app is a single page application which provides functionality to add, view, modify and delete addresses in a database through ``actionable'' web elements such as \emph{buttons} and \emph{links}, highlighted in their corresponding pages. 

\subsection{Automatic Test Generation for Web Apps}\label{sec:sec_bg_modelInference}

In this subsection, we describe automated model inference through state exploration and model-based test generation employed by existing techniques and their limitations.


\header{Model inference} \label{sec:modelInference_existing}
Automated model inference is an iterative process of exercising the functionality of a given web app by triggering  events on actionable elements~(\actionable{}), such as  \emph{button clicks}, and capturing the resulting state transitions (\transition{})  as a graph-based model (\model). Formally:
\change{}
\begin{definition}[\textbf{State Transition~(\transition{x})}]\label{def:def_transition}
is a tuple ($\mathcal{S}_{src}$, \actionable{x}, \StateVertex{tgt}) where exercising an actionable \actionable{x} in a state \StateVertex{src} produces a transition to state \StateVertex{tgt}.
\end{definition}
\stopchange{}
\begin{definition}[\textbf{Application Model~(\model)}]\label{def:def_model}
is a directed graph (\{\StateVertex{ 1}..\StateVertex{n}\},~ \{\transition{1}..\transition{m}\}) 
with \emph{app states}~(\StateVertex{a}) as nodes and \emph{state transitions}~(\transition{x}) as labelled directed edges between nodes. 
\end{definition}

  Current model inference techniques rely on a state abstraction function~($SAF$), that determines  similarity between two given  states $p_1$ and $p_2$ in order to avoid redundancies in the captured model and duplication of exploration effort. Formally:
  
\begin{definition}[\textbf{State Abstraction Function (${SAF}$)}]\label{saf}
is a pair ($dfunc$,~$t$), where $dfunc$ is a similarity function that computes the distance between any two given web pages $p_1$, $p_2$, and $t$ is a threshold defined over the output values of $dfunc$. $SAF$ determines whether the distance  between  $p_1$ and $p_2$ falls below $t$.

$$
   SAF(dfunc, p_1, p_2, t) \left\{
   {\arraycolsep=1.4pt
     \begin{array}{lll}
       true&:dfunc(p_1, p_2) < t \\
       &\\
       false&:otherwise\
     \end{array}
     }
   \right.
$$\end{definition}


 \begin{figure}[]
 
\centering
\includegraphics[width=0.8\columnwidth]{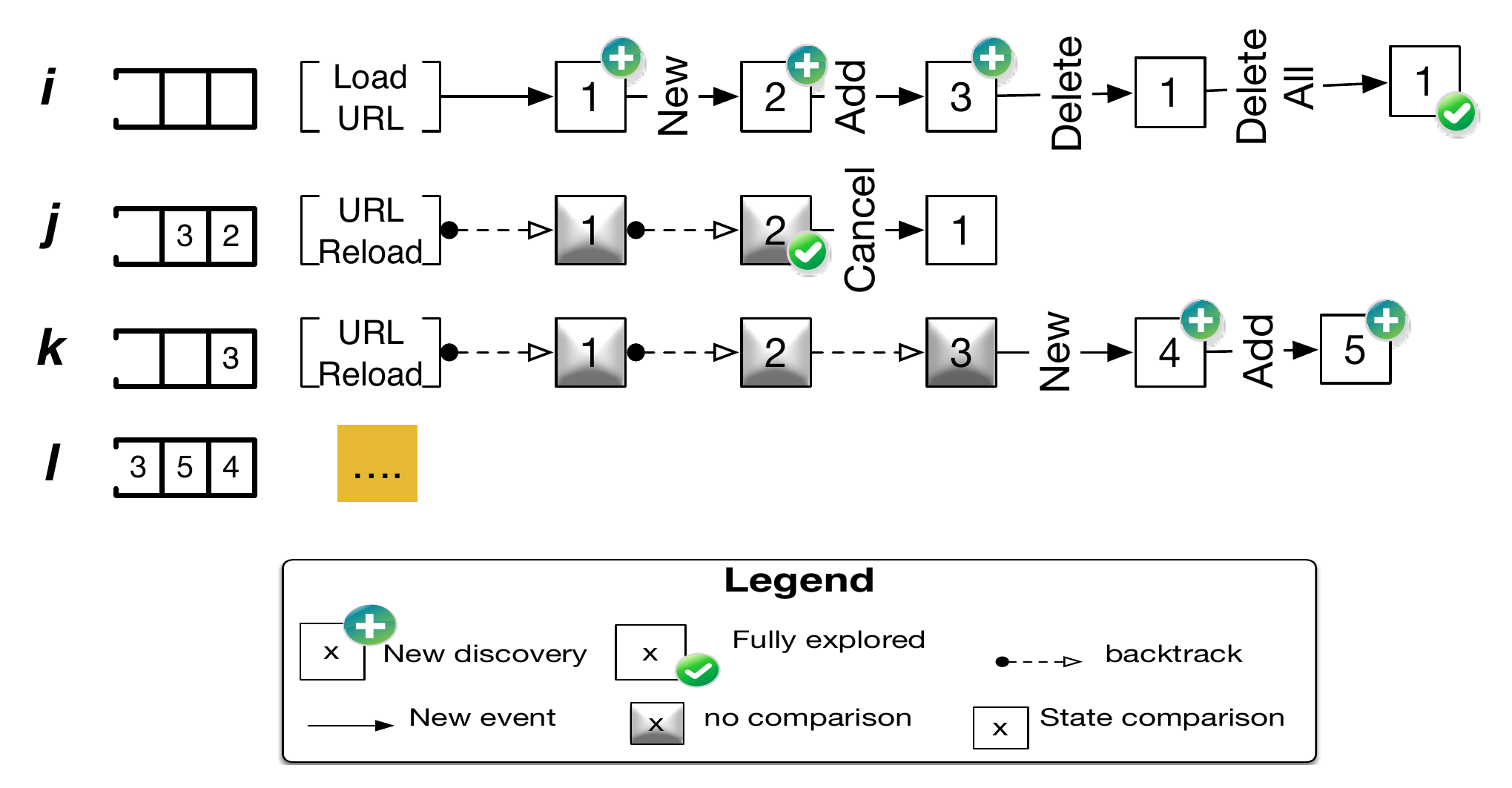}
\caption{Model inference for the motivating example.}
\label{fig:fig_crawlPaths}
\vspace{-0.5em}
\end{figure}

 \begin{figure}[]
\centering
\includegraphics[trim=0cm 0cm 0cm 0cm, clip=true, scale=0.38]{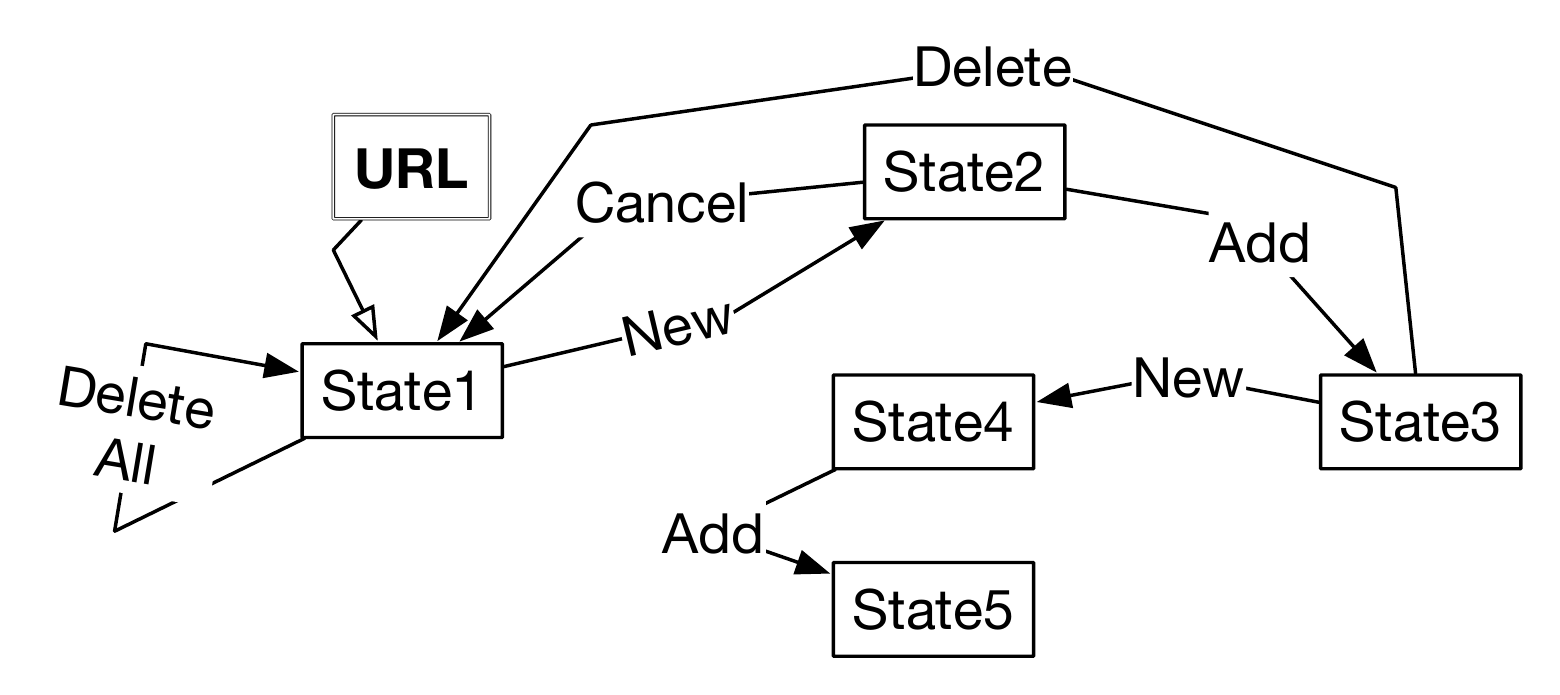}
\caption{Inferred model of the motivating example.}
\label{fig:fig_navModel}
\end{figure}
  \autoref{fig:fig_crawlPaths} illustrates the steps of model inference for our motivating app, assuming a depth-first exploration strategy is followed.
In the first iteration, labelled \emph{i}, the technique loads the app in the browser using its URL, and stores the corresponding state as the root node in the model.
Thereafter, each action performed on the web app can be either \emph{exploration} step or a \emph{back-tracking} step. 
 In each exploration step, depicted in solid arrows, an unexplored actionable~(\actionable{new}) from the current state is invoked and the resulting state is added to the model if it is
   deemed to be \emph{different} from every existing state in the model by a SAF as defined in defintion~\autoref{saf}.
   The observed state transition is then recorded as a directed edge between the source and target states in the model.  

 An iteration ends when the current state is fully explored and the next iteration starts by choosing an existing state in the model with unexplored actionables. 
 In order to reach the selected state, \emph{back-tracking} actions, indicated with broken arrows, are performed by using the transitions that are already recorded in the model.
%
  For example, iteration "{i}" ends upon reaching state \StateVertex{1}, which is fully explored at that point.
 The next iteration "{j}" then starts by choosing and navigating to state \StateVertex{2}, by using recorded transitions. Iteration "{l}" in~\autoref{fig:fig_crawlPaths} continues exploration by choosing one of the unexplored states \StateVertex{3}, \StateVertex{5} and \StateVertex{4}.
 \change
\header{Termination and stopping criteria}
Model inference techniques provide options to configure \emph{stopping criteria} such as exploration time limit in order to end the inference process. \emph{Termination} on the other hand happens when the technique decides that no unexplored actionables are left to exercise.  
\stopchange{}

\autoref{fig:fig_navModel} shows the model inferred for our motivating example at the end of iteration "{k}". 
\change
The model inference has not \emph{terminated} at this point because there are unexplored actionables available.  
   \begin{definition}[\textbf{Path~( \Path{} )}]\label{def:path}
A sequence of transitions~(\transition{0}...\transition{n}) is a \Path{} if for 0<=$i$<n, 
 \transition{i},  \transition{i+1}  $\in$ \Path{} $\implies$ \transition{i}(\StateVertex{tgt})~=~\transition{i+1}(\StateVertex{src}). 
\end{definition}
 \stopchange{}
 \header{Test generation} 
Once a model of web app is available, model-based test generation   
provides a set of paths~\{\Path{1}, \Path{m}\}, with adequate coverage of states and/or transitions in the model, where, each path~\Path{}~(formally defined in~\autoref{def:path}) is a sequence of recorded transitions. 
\autoref{listing:testCase} shows a test case \testCase{} generated from the inferred model of \autoref{fig:fig_navModel}. Each test case starts by loading the URL of the app and verifying the browser state to be the \StateVertex[0]{src}, i.e. the source state of the first transition of the path. Thereafter, for each transition~(\transition{x}), a \emph{test action} is derived from the actionable~\actionable{x} and a \emph{test oracle} is added for the target state~\StateVertex[x]{tgt}. The test case in \autoref{listing:testCase} is examining the path \Path{} : [\StateVertex{1}, \actionable{List}, \StateVertex{2}, \actionable{new}, \StateVertex{3}, \actionable{add}, \StateVertex{4}].

\begin{figure}
\begin{lstlisting}[language=Python, label={listing:testCase}, caption=Generated test case]
def Test1(): 
		driver.loadURL("Base_URL")
		assert(isEqual(driver.currentState, state1)
		driver.findElement("List").click()
		assert(isEqual(driver.currentState, state2)
		driver.findElement("New").click()
		assert(isEqual(driver.currentState, state3)
		driver.findElement("add").click()
		assert(isEqual(driver.currentState, state4)
\end{lstlisting}
\vspace{-1em}
\end{figure}   
\subsection{Automatic Test Generation Challenges}\label{sec:sec_challenges}
The main challenge in model inference of 
 web apps is the 
presence of a large number of nearly identical or near-duplicate web pages, 
which the existing techniques cannot identify effectively and as a result generate sub-optimal models.

While the presence of near-duplicates in the inferred web app model leads to generation of redundant test cases, it is also indicative of wasted exploration effort that could be spent discovering unseen states.
Typically, exploration of similar actions in near-duplicate states does not improve functional coverage of the model but instead can lead to creation of even more near-duplicates, as can be seen even in our simple example app.
In our example, the states \StateVertex{3}, \StateVertex{5} and \StateVertex{6} shown in~\autoref{fig:fig_near-duplicates} are all considered \emph{functional near-duplicates} as they all offer similar web app functionality, and model inference may never \emph{terminate} if similar actions \actionable{New} and \actionable{Add} are explored in each near-duplicate state.
 

In our previous work, we proposed~\cite{ndStudy2020} to categorize a given pair of web pages as either clone ($Cl$), near-duplicate~($Nd$) or distinct~($Di$) by labelling the observed changes between them. 
We formally defined \emph{Near-duplicates} in web testing as:
  \begin{definition}[\textbf{Functional Near-Duplicate (Nd)}]\label{def:nd}
  A given state-pair ($p_1$, $p_2$), is considered to be a functional near-duplicate if the \emph{changes} between the states do not alter the overall functionality of either state.

Near-duplicates are further divided in three categories based on the nature of changes between the two pages:

  \begin{itemize}
  \item Cosmetic~(\ndOne): Changes such as different advertisements, that are irrelevant to functionality of web app. 
  \item Dynamic Data (\ndTwo): Changes are limited to data while the page structure remains the same. Example \emph{(\StateVertex{3}, \StateVertex{6})}.
  \item Duplication (\ndThree): Addition or removal of web elements equivalent to existing web elements. { Example \emph{(\StateVertex{3}, \StateVertex{5})}}.
  \end{itemize}
  \end{definition}

Researchers have employed various similarity functions and abstractions for web pages based on their DOM tree-structures~\cite{Mesbah:2009, WESS01a}, and visual screenshots~\cite{Choudhary-2012-ICST,   Choudhary-2013-ICSE, Mahajan-2015-ICST, Mahajan-2014-ASE} to perform state comparison. 
However, a study~\cite{ndStudy2020} on near-duplicates in web testing shows that the whole-page based SAFs currently being used by existing model inference techniques cannot reliably identify near-duplicates and as a result infer imprecise and incomplete models.


\begin{table}[]
\caption{Raw distances of state-pairs}
\label{table:table_motivation}
\footnotesize
\centering

\resizebox{0.95\columnwidth}{!}{

\setlength{\tabcolsep}{4pt}     
\renewcommand{\arraystretch}{1.0}

\begin{tabular}{lrrrr@{\hskip 0.3cm}rrrrrrr}

\toprule
\textit{\textbf{~~~\rot{state-pair}}} & \rot{\textit{\textbf{RTED~\cite{Pawlik:2015:ECT:2751312.2699485}}}} & \rot{\textit{\textbf{Levenshtein~\cite{levenshtein1966bcc}}}} & \rot{\textit{\textbf{TLSH~\cite{tlsh}}}} & \rot{\textit{\textbf{SimHash~\cite{Charikar:2002:SET:509907.509965}}}} & \rot{\textit{\textbf{HYST~\cite{Swain}}}} & \rot{\textit{\textbf{BlockHash~\cite{Yang-blockhash}}}}& \rot{\textit{\textbf{PHASH~\cite{phash}}}}  & \rot{\textit{\textbf{PDIFF~\cite{Yee:2001:SSV:383745.383748}}}} & \rot{\textit{\textbf{SIFT~\cite{Lowe1999}}}} & \rot{\textit{\textbf{SSIM~\cite{1284395}}}} & \rot{\textit{\textbf{Human}}} \\ 
\cmidrule(l{1em} r{1em}){2-5}
\cmidrule(l{1em}){6-11}
&\multicolumn{4}{c}{DOM} & \multicolumn{6}{c}{VISUAL} &\\
\midrule
(\StateVertex{3}, \StateVertex{6}) & 0.0 & 0.002 &  1 & 0 & 91 & 0 & 0 & 0.0006 & 5 & 0.003 & $Nd_2$ \\

(\StateVertex{1}, \StateVertex{3})& 0.13 & 0.08 & 36 & 0 & 9751 & 4 & 2 & 0.008 & 14 & 0.04 & $Di$ \\


%
(\StateVertex{3}, \StateVertex{5}) & {0.16} & {0.12} & {106} & 0 & {27683} & {5} & {2}  & {0.045} &{14} & {0.05} & $Nd_3$ \\

%
%
 \specialrule{\heavyrulewidth}{4pt}{4pt}
\end{tabular}

}

\end{table}

To illustrate the limitations of the SAFs currently being used, we analyzed three state-pairs of our example app  in~\autoref{table:table_motivation}. 
%
For each of the three state-pairs, the table shows the distance between states computed by each of the ten state abstraction techniques and the \emph{human} classification process followed in the study~\cite{ndStudy2020}.
 
 Consider the states \StateVertex{1}, \StateVertex{3}, \StateVertex{5}, and \StateVertex{6} from \autoref{fig:fig_near-duplicates}, all of which display stored addresses. 
When examined manually, \StateVertex{1} is considered distinct~($Di$) from \StateVertex{3}, \StateVertex{5} and \StateVertex{6} as it does not contain table row functionality to select an address entry.  
 However, \StateVertex{3} and \StateVertex{6} differ only by the \emph{data} in a table row, which does not alter the functionality and hence they are considered as functional near-duplicates. Further, even-though \StateVertex{5}  contains extra table rows, they only duplicate the functionality of a single row in \StateVertex{3}. Therefore, a human tester would label state-pairs such as (\StateVertex{3}, \StateVertex{5}) to be functional near-duplicates. 
 
The 10 comparison techniques, as shown in the \autoref{table:table_motivation}, consider the state-pair (\StateVertex{3}, \StateVertex{5}) to be the farthest apart of the four state-pairs even though they are functionally equivalent. 
If the distance thresholds were set to be higher, states such as \StateVertex{3} might be  discarded from the model as they will be considered equivalent to \StateVertex{1}, making the model \emph{incomplete}. On the other hand,  lower thresholds would make the model imprecise with the presence of states such as the state \StateVertex{5}. 
This shows that threshold-based whole-page comparison cannot produce optimal web app models.

\header{Test breakages}
In addition to model inference, similarity between two given web page states is crucial in 
 generating effective test oracles as well.
If the state comparison techniques are sensitive to minor changes unrelated to functionality in modern apps, test oracles can break and  result in a high number of false positive test failures. False positive test failures necessitate costly manual analysis and impact the effectiveness of automated regression testing techniques. 

\change
We identified duplicated functionality within a web page to be the root cause of \ndThree near-duplicates that make existing model inference and test generation techniques impractical for modern web apps. 
This observation that ``whole-page techniques'' cannot detect \ndThree near-duplicates 
motivated us to 
investigate the idea of decomposing a given web page into smaller fragments for a more fine-grained analysis.

\subsection{Page Fragmentation}\label{sec:sec_bg_fragmentation}

Page fragmentation, also known as page segmentation, is the decomposition of a given web page into smaller fragments or segments.
In existing research, the most popular downstream tasks such as content extraction that apply page fragmentation relate to human consumption of web pages and focus on extracting textual semantics.

The "VIsion-based Page Segmentation algorithm" (VIPS)~\cite{vips}, proposed in 2003, is the de-facto standard for web page segmentation. A recent large scale empirical study~\cite{fragmentationStudy} compares five page segmentation techniques using a dataset of 8,490 web pages and concludes that VIPS is still the overall best option for page fragmentation.
VIPS employs a top-down approach, where, for each HTML node in the DOM~\cite{DOM}, a DoC (Degree of Coherence) is assigned to indicate coherence of the content in the block based on visual perception. Then it tries to find the separators between these extracted blocks. Here, separators denote the horizontal or vertical lines in a webpage that visually cross with no blocks. Finally, based on these separators, it extracts a hierarchy of fragments.  Each extracted fragment visually conforms to a rectangle in the page between two horizontal and vertical separators. All the DOM nodes that are part of the rectangle are then considered to be part of the fragment. We refer the interested reader to the original VIPS paper for more details of the algorithm~\cite{vips}.



 In the next section, we explain how we leverage page fragmentation to overcome existing challenges in model inference and regression test generation for modern web apps. 
\stopchange

\section{Approach}
At a high level, our approach, called \toolname, relies on the insight that a web page is not a singular functional entity, but a set of functionalities, where each functionality may be available in more than one page. Based on this insight, we propose a novel state abstraction that defines a page as a \emph{hierarchy of fragments}, where each fragment represents a semantic sub functionality. Our model inference technique then employs this state abstraction to detect near-duplicates and optimize the state space exploration by prioritizing actions that belong to unique page fragments in each page. From the inferred model, we subsequently generate test cases with robust assertions that rely on our fragment-based abstraction and are capable of reporting warnings in addition to errors by utilizing knowledge gained during model inference. Next, we describe our state abstraction, followed by our model inference, and test generation techniques.

\begin{figure}
\centering
\includegraphics[trim=0cm 0cm 0cm 0cm, clip=true, width=0.98\columnwidth, height=3.2in]{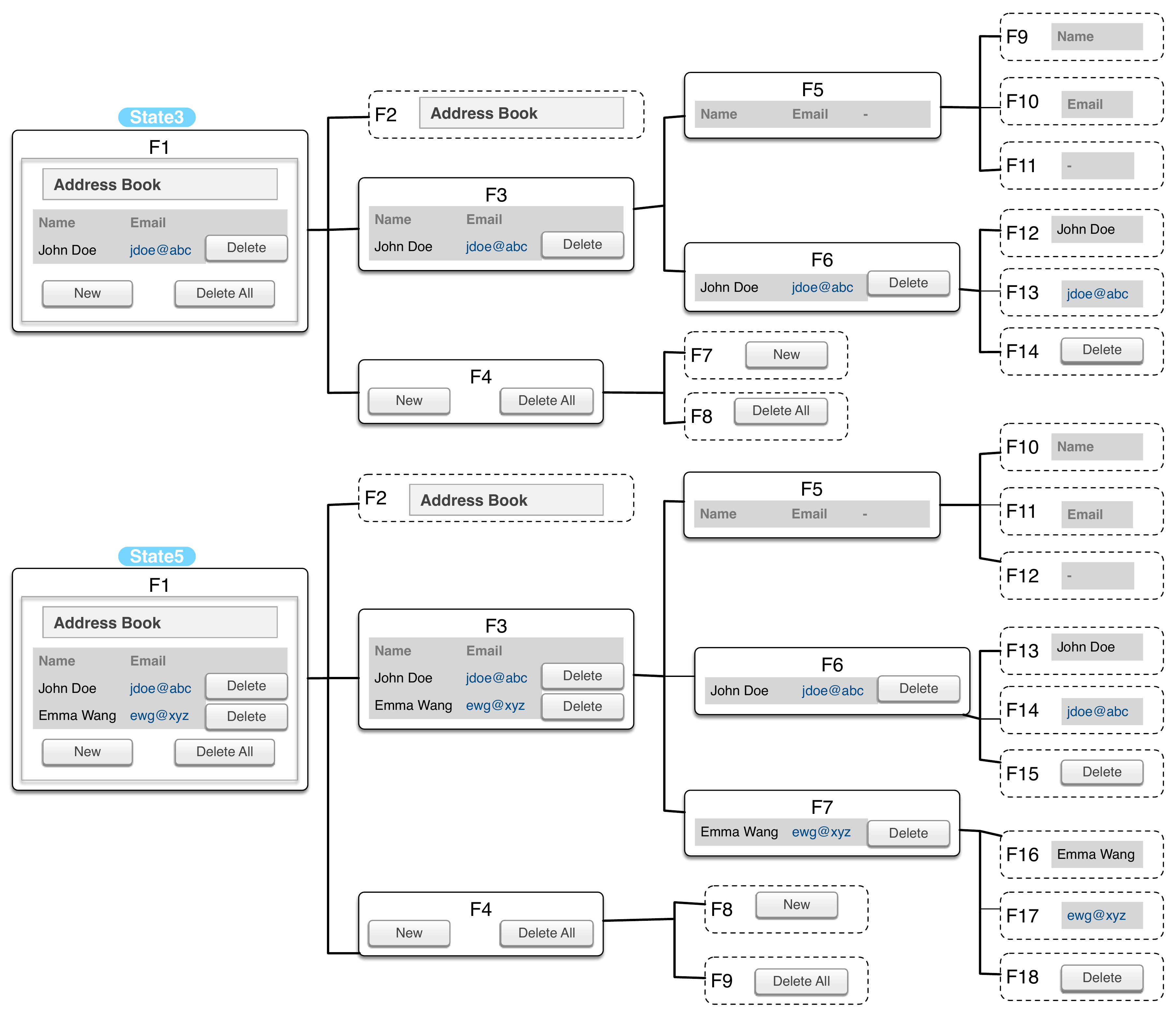}
\caption{Fragment-based state comparison in \toolname}
\label{fig:fig_fragmentation}
\end{figure}

\subsection{Fragment-based State Abstraction}

We consider a web page or a state as a hierarchy of fragments, 
 where, each fragment is a portion of the state and represents functionalities offered by its child fragments. 

\begin{definition}[\textbf{Application State~$(\mathcal{S}) $}] \label{def:appState}is a tuple ($\mathcal{D}$, $\mathcal{V}$, \Fragment{root}) where $\mathcal{D}$ is the dynamic DOM\change~\cite{DOM}\stopchange{} of the page, $\mathcal{V}$ is the screenshot of the page and \Fragment{root} is the root fragment.
\end{definition}

The root fragment,  \Fragment{root}, of a state $\mathcal{S}$ is the full page, and has no parent. It has all the nodes in the DOM tree $\mathcal{D}$ of $\mathcal{S}$. Therefore, comparing two states is the same as comparing their root fragments.

\begin{definition}[\textbf{Fragment~($\mathcal{F}$)}]\label{def:fragment}
is a tuple ($\mathcal{N}$, $V$, $\{ \mbox{\Fragment{c_0}}, \mbox{\Fragment{c_1}} \dots \}$) where $\mathcal{N} \mbox{ is the set of DOM nodes of }  \mathcal{F}$, 
 $V$ is the screenshot of  $\mathcal{F}$ and 
each \Fragment{c_i} is a child fragment.
 \end{definition}






\begin{algorithm}[t]
{\scriptsize
\SetAlgoLined
\DontPrintSemicolon
\SetNoFillComment

    \SetKwFunction{FMain}{Classify}
    \SetKwProg{Fn}{Function}{:}{}
    
    \Fn{\FMain{\Fragment{1}, \Fragment{2}}}{
    \tcp*[l]{$clone(Cl),distinct(Di)$,\ndTwo,\ndThree}    

\KwOut{$class$}
        $ \mathcal{N}_{diff}  \longleftarrow $ treediff ( \Fragment{1}.$\mathcal{N}$, \Fragment{2}.$\mathcal{N}$ ) \;    
        
        \uIf(\tcc*[f]{Matching DOM}){$ \mathcal{N}_{diff}  = \phi$}
	{
	  	$ \mathcal{V}_{diff}  \longleftarrow $ imagediff ( \Fragment{1}.$V$, \Fragment{2}.$V$ ) \; 
		\eIf(\tcc*[f]{ Matching Screenshots}){$\mathcal{V}_{diff} = \phi$ }
			{\textbf{return} $Cl$  \tcc*[f]{ class: $clone$}} 
			{\textbf{return} \ndTwo \tcc*[f]{class: \ndTwo}}
	}
	\Else(\tcc*[f]{ Check Child Fragments?}){ 
		\textbf{return} MapChildFragments($\mathcal{N}_{diff}, $\Fragment{1}, \Fragment{2})\;
	}       
}
\textbf{End Function}

\vspace{1.5em} 
 \SetKwFunction{FMain}{MapChildFragments}
 \SetKwProg{Fn}{Function}{:}{}

   \Fn{\FMain{$\mathcal{N}_{diff}, $\Fragment{1}, \Fragment{2}}}{
   \tcp*[l]{ $distinct (Di)$, \ndThree}
  \KwOut{$class$} 
\tcc{Mapping child fragments for every changed node}
		 \ForEach(){$ n \in \mathcal{N}_{diff} $} 
	       	{
			\emph{found} $\longleftarrow$ false    \;
			\Fragment{oth}   $\longleftarrow$  (~$n \in $ \Fragment{1}.$\mathcal{N}$) $?$  \Fragment{2} $\colon$  \Fragment{1} \;

			\Fragment{clo} $\longleftarrow closest$~(~n~) \;
			
			 \ForEach{ \Fragment{chi} $\in$ \Fragment{oth}.childen}
	       			{
					  \If{Classify~(~\Fragment{clo}, \Fragment{chi}~)~ $!=$ ~$Di$}
						{
	  						\emph{found} $\gets$ true  \tcc*[f]{Found mapping}
 						}
	        			}
			\If(\tcc*[f]{No mapping found}){ ! \emph{found}} 
			{\textbf{return} \emph{Di}  \tcc*[f]{class: $Distinct$}}
		}
		\textbf{return} \ndThree  \tcc*[f]{class: \ndThree}
}
	\textbf{End Function}	
\caption{\normalsize \textbf{ Fragment-based classification}}
\label{algo:algo_fragComp}
}
\end{algorithm}

\header{Comparing fragments}
 Using this fragment-based representation, we classify a given state-pair by comparing the fragments they are composed of. ~\autoref{fig:fig_fragmentation} shows the fragment hierarchies for states  \StateVertex{3} and  \StateVertex{5} from our motivating example in
\autoref{fig:fig_near-duplicates}. 
  Our fragment-based classification, which takes both structural and visual aspects into account is shown in Algorithm~\ref{algo:algo_fragComp}. The structural aspect\change~(line 2)\stopchange{} uses the nodes on the DOM subtree of the fragment after pruning textual content and attributes. The visual aspect\change~(line 4)\stopchange{} uses a localized screenshot of the fragment.
  
As \autoref{algo:algo_fragComp} shows, 
we combine structural and visual analysis to identify near-duplicates.
Using visual similarity instead of element attributes and textual data from DOM allows us to disregard changes in the DOM that have no visual impact. In addition, we found that visual comparison is effective in identifying changes in dynamic web elements such as carousels that use only JavaScript and CSS.
 

\toolname classifies two fragments to be \emph{clones} if their structural and visual properties are exactly the same. We employ APTED~\cite{apted} and Color Histogram~\cite{Swain} to compare the structural and visual aspects of the fragments,  respectively. These techniques have been employed individually as state abstractions for whole web pages in the literature~\cite{Choudhary-2013-ICSE, ndStudy2020}, \change but have not been combined to determine state equivalence.\stopchange{} 
When the fragments differ visually, but are found to be structurally equivalent, \toolname considers the fragments to be near-duplicates of the type \ndTwo (see Definition \autoref{def:nd}).

 In case they differ structurally, as root fragments of \StateVertex{3} and \StateVertex{5} do for example shown in~\autoref{fig:fig_fragmentation}, the classification is performed  by a mapping function \change(lines 14-28)\stopchange{} that extracts all changed DOM nodes between the two fragments and maps the corresponding fragments in the hierarchy. 	
 
 \change
Given a pair of fragments \Fragment{1} and \Fragment{2} and a list of changed nodes between the two fragments $\mathcal{N}_{diff}$, the function $MapChildFragments$ gets the closest fragment \Fragment{clo}  for each changed DOM node~($n$). The closest fragment for a DOM node is the smallest child fragment in the fragment hierarchy containing $n$. 
Thereafter, in lines 19-23, it attempts to find an equivalent fragment for the closest fragment in the fragment hierarchy of the other fragment. 
For example, if $n_1$ is a changed node that belongs to \Fragment{1}, with \Fragment{clo} as its closest fragment, then \Fragment{oth} is \Fragment{2} and we attempt to find if any of the child fragments of \Fragment{2} are equivalent to \Fragment{clo}.
 If the closest fragment for any of the changed DOM nodes cannot be mapped to a child fragment of the other fragment~(lines 24-26), we declare the two given fragments \Fragment{1} and \Fragment{2} to be distinct~($Di$). Otherwise, the two fragments are considered to be near-duplicates of type \ndThree. 


 In the example shown in~\autoref{fig:fig_fragmentation}, when classifying the root fragments for \StateVertex{3} and \StateVertex{5}, $MapChildFragments$ would be called to classify \Fragment{3} of states \StateVertex{5} and \StateVertex{3}. 
 All the changed DOM nodes~($\mathcal{N}_{diff}$) between the two fragments belong to \Fragment{7} in  \StateVertex{5},
 As the function iterates~(line 19) through child fragments to find a mapping, eventually \Fragment{7} of \StateVertex{5} will be found to be equivalent to \Fragment{6} of \StateVertex{3}. 
 As a result, the fragment pair would be classified as \ndThree. 
 As the rest of the fragments are equivalent, the overall state-pair (\StateVertex{3}, \StateVertex{5}) would be classified as \ndThree near-duplicates as well. 
\stopchange{}

 When the fragments containing changed DOM nodes cannot be mapped, \toolname considers the two states to be \emph{distinct}.
 One such example is the state-pair (\StateVertex{1}, \StateVertex{3}), which is classified to be \emph{distinct} because the table row in \StateVertex{3} contains no equivalent fragment in \StateVertex{1}. 
 On the other hand, (\StateVertex{3}, \StateVertex{6}) are classified as \ndTwo because the abstracted DOM hierarchy is equal, making them structurally similar, while the data changes inside the table make them visually dissimilar. 

\subsection{Fragment-based Model Inference}\label{sec:modelInference}



\toolname infers the model of a given web app by iteratively triggering user interactions on actionables and recording the corresponding state transitions much like the existing techniques~(described in \autoref{sec:modelInference_existing}). 
\change
After an action~(\actionable{}) is performed on a source state, the resulting browser state is compared to all the existing states in the model using the $Classify$ function~\autoref{algo:algo_fragComp}. If the classification is clone or 
\ndTwo
for any of the existing states, the new state is discarded, otherwise, it is added to the model. 

However, existing techniques assign equal importance to every actionable of a newly discovered state, often wasting exploration effort exercising similar actionables.\stopchange{}
In contrast, \toolname identifies similar actionables through fragment analysis to diversify the exploration.  In this section, we describe how our fragment analysis is used to 1) diversify state exploration to discover unique states faster, and 2) identify \emph{data-fluid}  fragments to generate effective test oracles. 

\subsubsection{Exploration strategy} \label{sec:explorationStrategy}
\toolname ranks the actionables and states using \autoref{eq:eq_actionScore} and \autoref{eq:eq_stateScore}, respectively, using the fragment comparisons in~\autoref{eq:eq_equivActions} to determine the equivalence of actionables.
 Actionables are special DOM nodes such as buttons which are used for user interaction.

{\small Where a function $Xpath$(\actionable{}, \Fragment{}) provides a relative XPath \change expression~\cite{xpath}\stopchange{}  for \actionable{} within \Fragment{}, we decide the equivalence of two actionables~\actionable{x} $\in$ \Fragment{x}, \actionable{y} $\in$ \Fragment{y} using:}
\begin{multline}
\label{eq:eq_equivActions}
	\text{\actionable{x}} \equiv  \text{\actionable{y}} \Longleftrightarrow 
	\text{\footnotesize	(Classify(\Fragment{x}, \Fragment{y}) = Clone $\vee$ \ndTwo)} \\ 
	\wedge 
	 \text{\footnotesize($Xpath$(\actionable{x}, \Fragment{x}) == $Xpath$(\actionable{y}, \Fragment{y}))}
\end{multline}
 
\noindent{\small Given a constant $c_0$ where $0< c_0 <= 1$, and \actionable{eq} is an equivalent actionable determined using ~\autoref{eq:eq_equivActions}, the score for an actionable \actionable{} is computed as:}
\begin{equation}
\footnotesize
\label{eq:eq_actionScore}
 score(\text{\actionable{}})=\begin{cases}
    -1, & \text{if \actionable{}.explored}.\\
    0, & \text{if } \exists \text{ \actionable{eq}} \mid \text{\actionable{eq}.explored} .\\
    c_0 * size(\{\text{\actionable{eq}} \dots \}) , & \text{otherwise}.
  \end{cases}
\end{equation}
\vspace{-0.3em}

\noindent{\small The score for a state \StateVertex{} is computed using the scores of actionables in the state as:}
\begin{equation}
\footnotesize
\label{eq:eq_stateScore}
	score(\text{\StateVertex{}}) =  \sum_{n=1}^{\text{size(\{\actionable{}\})}} \mbox{score(\actionable{})}
\end{equation}


Using the \autoref{eq:eq_stateScore}, \toolname chooses the next state to explore based on the total score of the actionables in each state.
The score for each actionable~(\actionable{}) is assigned by \autoref{eq:eq_actionScore} based on its equivalent actionables~(\actionable{eq}) across states. Once a state is chosen, again the actionable with a higher score is chosen for exploration, which helps prioritize unexplored actionables that have high repetition such as navigation links. 
Equivalence of actionables is established using the fragments that contain them using the~\autoref{eq:eq_equivActions}.
Through these equations, every time an actionable is explored, \toolname de-prioritizes all of its equivalent actionables to diversify the exploration.  \change As a result of~\autoref{eq:eq_equivActions}, this de-prioritization often reduces redundant exploration effort spent in already seen functionality. \stopchange{}

\begin{figure}[]
\centering
\includegraphics[trim=0cm 0cm 0cm 0cm, clip=true, width=0.75\columnwidth]{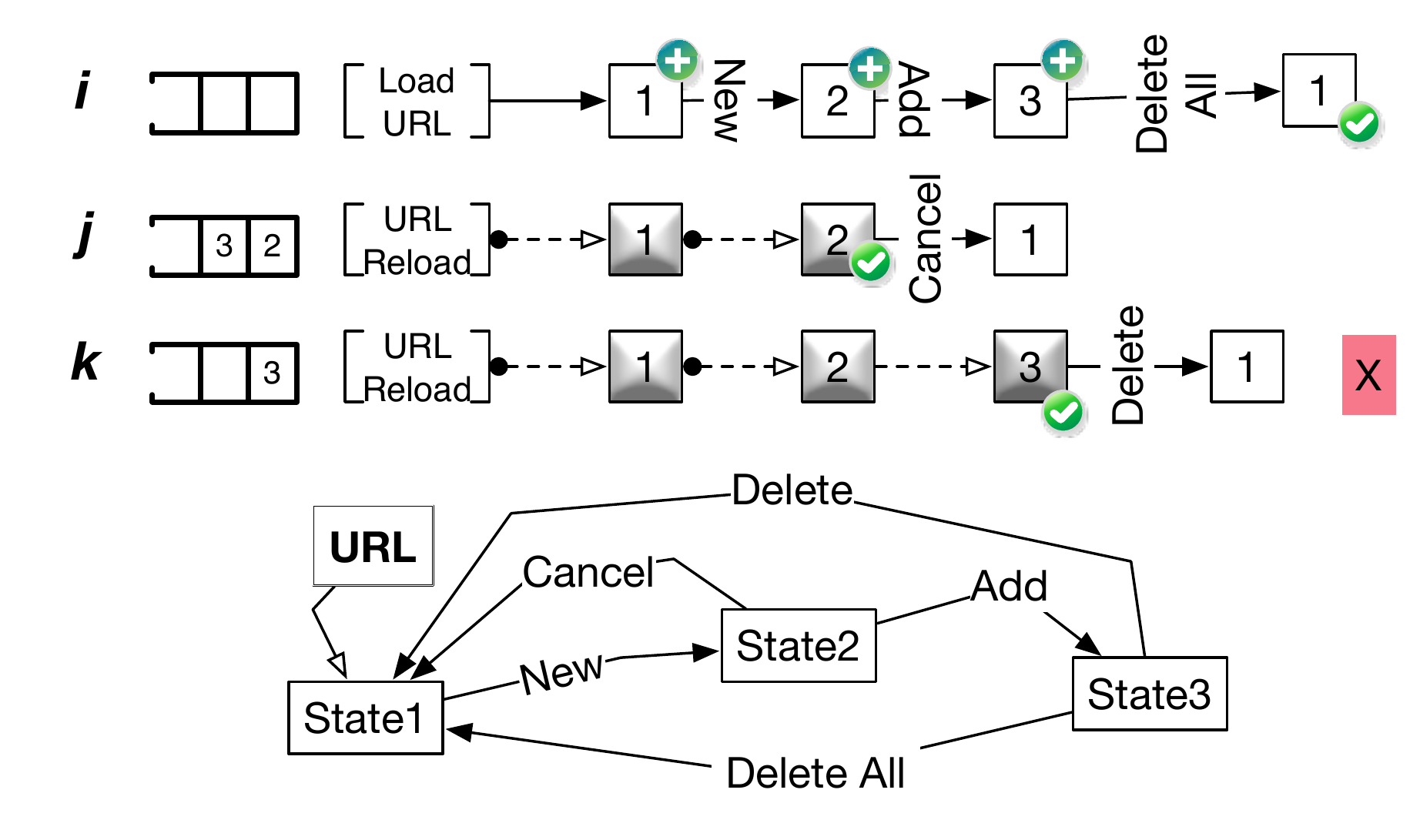}
\caption{Model inferred by \toolname}
\label{fig:fig_navModel_fraggen}
\vspace{-0.5em}
\end{figure}

\autoref{fig:fig_navModel_fraggen} shows the model inference performed by \toolname for our running example. As it can be seen, \toolname can generate a more 
precise model, exercising actionables that are considered unique by our fine-grained fragment analysis. For example, \toolname can identify \actionable{New} in \StateVertex{3} to be equivalent to that of \StateVertex{1} and doesn't consider it to be unexplored, thus \emph{terminating} the exploration. On the other hand, as shown in~\autoref{fig:fig_navModel}, existing techniques exercise \actionable{New} on \StateVertex{3} as well, leading to the creation of \StateVertex{4} and \StateVertex{5} and necessitating further exploration of the same actionables in the newly added states as well. \change In fact, without identifying the equivalence of such repeated actionables, state exploration will never \emph{terminate} even for this simple example app we present in the paper. Existing techniques would require a stopping criterion such as a time limit to end the inference process and will likely generate an imprecise app model with a high number of near-duplicates. \stopchange{}
 
 \change
\header{Configuration options}
\emph{constant $c_0$ in ~\autoref{eq:eq_actionScore}}: $c_0$ determines the weight given to the presence of duplicates for an action that is yet to be explored. In our experiments, we used the default value which is set to~1. We provide the ability to tune it if necessary for specific web applications. 
The value of constant c0 in~\autoref{eq:eq_actionScore} can impact prioritization of states, where, a high value could delay exploration of a potentially more interesting newly discovered state. This scenario is possible if the number of duplicates for a single unexplored actionable outnumber the total actionables in a newly discovered state. In this scenario, the older states containing this particular unexplored action would get a higher priority even if a majority of the actions in the state itself are already explored.
However, once any one of the instances of the particular actionable is exercised, the value of constant $c_0$ becomes irrelevant.

\emph{Termination}:\stopchange{}
We designed \toolname to be configured to limit the amount of duplication in order to avoid generating inflated models. By default, \toolname does not exercise any actionable which is found to be similar to an already explored actionable. It will \emph{terminate} once all unique actionables are exercised.
However, this duplication in exploration effort can sometimes be necessary to fully explore app functionality. 
Therefore, we provide configuration to continue exploration using unexplored actionables that were previously skipped because they were similar to already explored actionables. 
\toolname would continue to use ~\autoref{eq:eq_actionScore} and \autoref{eq:eq_stateScore}, for prioritization.

\change
\subsubsection{Memoization and data-fluid fragments} \label{sec:memoization}
During the model inference process, recall that after every exploration step, \toolname decides if the resulting browser state should be retained in the model by comparing it to the existing states using the $Classify$ function. Our exploration strategy also relies heavily on the $Classify$ function to search for diverse actions as seen in~\autoref{eq:eq_equivActions}.

From~\autoref{algo:algo_fragComp}, it can be seen that $Classify$ for a given pair of fragments is recursive in nature that may require comparing the corresponding child fragments.  
To improve its performance, we apply a classic technique for recursive algorithms called memoization, which aims to avoid recalculation of results for sub-problems. 
Here, instead of storing just the comparison results for every pair of fragments, 
we employ a map-based implementation with unique fragments as map keys and a list of their duplicates as the values.

Whenever a new state is added to the model, we update the map by comparing each of the fragments in the new state to existing unique fragments. If any of the fragments in the new state is not a clone ($Cl$) of existing unique fragments, it is added to the map as a new unique fragment. Otherwise, it is added as a duplicate of the unique fragment it is found to be a clone of.

Further, we utilize this memoization technique using the map of fragments to identify \emph{data-fluid} fragments. A fragment is data-fluid if it is found to be an \ndTwo near-duplicate of at-least one other fragment in the map. Intuitively, we are trying to recognize the parts of the dynamic DOM, specific to the web app being tested, that are likely to have data changes. 
Any clone of a data-fluid fragment is also data-fluid. For example, in ~\autoref{fig:fig_fragmentation}, \Fragment{6} in \StateVertex{3} and \Fragment{6}, \Fragment{7} in \StateVertex{5} are identified to be \emph{data-fluid}. 

In practice, there could be multiple causes for changing data in fragments; such as a DOM element that displays current time.
Indeed, this ability to mutate dynamic DOM in real-time is the primary reason that developers are able to create highly interactive and responsive web apps.
However, such changes also pose a challenge to web testing in the form of fragility of the automatically generated test oracles which fail in response to every change in the dynamic DOM.
 In the next section, we describe how \toolname mitigates this challenge by making use of memoization and data-fluid fragments to generate robust test oracles.
\stopchange{}







\subsection{Test Generation}\label{sec:sec_testGeneration}

UI Test cases, such as the one shown in ~\autoref{listing:testCase}, are sequences of UI events derived  from inferred web app models as discussed in~\autoref{sec:modelInference_existing}.
We designed \toolname to generate tests using \emph{exploration paths} 
which are essentially inferred model iterations as shown in~\autoref{fig:fig_navModel_fraggen}, and contain at least one unexplored action in each new path.
\change
Recall~(from~\autoref{sec:modelInference_existing}) that a model iteration starts by reloading the URL and ends when the browser state does not have any unexplored actions remaining. 
\stopchange{}
As the exploration paths or model iterations cover all states and transitions, test cases generated by \toolname also cover all the states and transitions in the model. 


\change
Each exploration path translates to a test case that starts with loading URL, similar to the test case shown in~\autoref{listing:testCase}. For each transition~(\transition{} -> (\StateVertex{src}, \actionable{}, \StateVertex{tgt})) in an exploration path~(\Path{}, definition~\autoref{def:path}), we generate a test step to perform the action~(\actionable{}) and generate assertions that compare corresponding recorded model states with the browser states by using the function $Classify$~(\autoref{algo:algo_fragComp}).

In the remainder of this section, we describe the challenges in generating robust test assertions and how \toolname utilizes fragment analysis to tackle this challenge.\stopchange{}

\subsubsection{Test assertions}
An important but often neglected aspect of regression testing is the creation of effective 
 test assertions or test oracles, which are key in detecting app regressions. 
 Existing test generation techniques generate fragile/brittle test assertions~\cite{torsel_automated_2011},
which results in many false positive test failures. 
 

 Generating web test assertions is considered difficult~\cite{oracle_survey}, primarily because they require a reliable state comparison to handle near-duplicates.
Previously, researchers have relied on manual specification of DOM invariants~\cite{mesbah:tse12} specific to a web app. 
 Although effective, such a procedure is not scalable for complex large applications and requires significant human effort as well as domain knowledge.

Our test assertions use the fragment-based state abstraction to identify changes between recorded states in the model and the states observed during test execution. 
We designed \toolname's test execution to utilize
 data-fluid fragments identified during model inference to assign importance to the changes detected by our state comparison.
As a result, \toolname can produce fine-grained warnings with different levels of severity based on the characteristics of detected changes. 
Test failures can then be declared based on the severity of warnings, reducing false positive test failures or test breakages caused by unimportant changes. 
\begin{figure}[]
\centering

\includegraphics[width=0.85\linewidth]{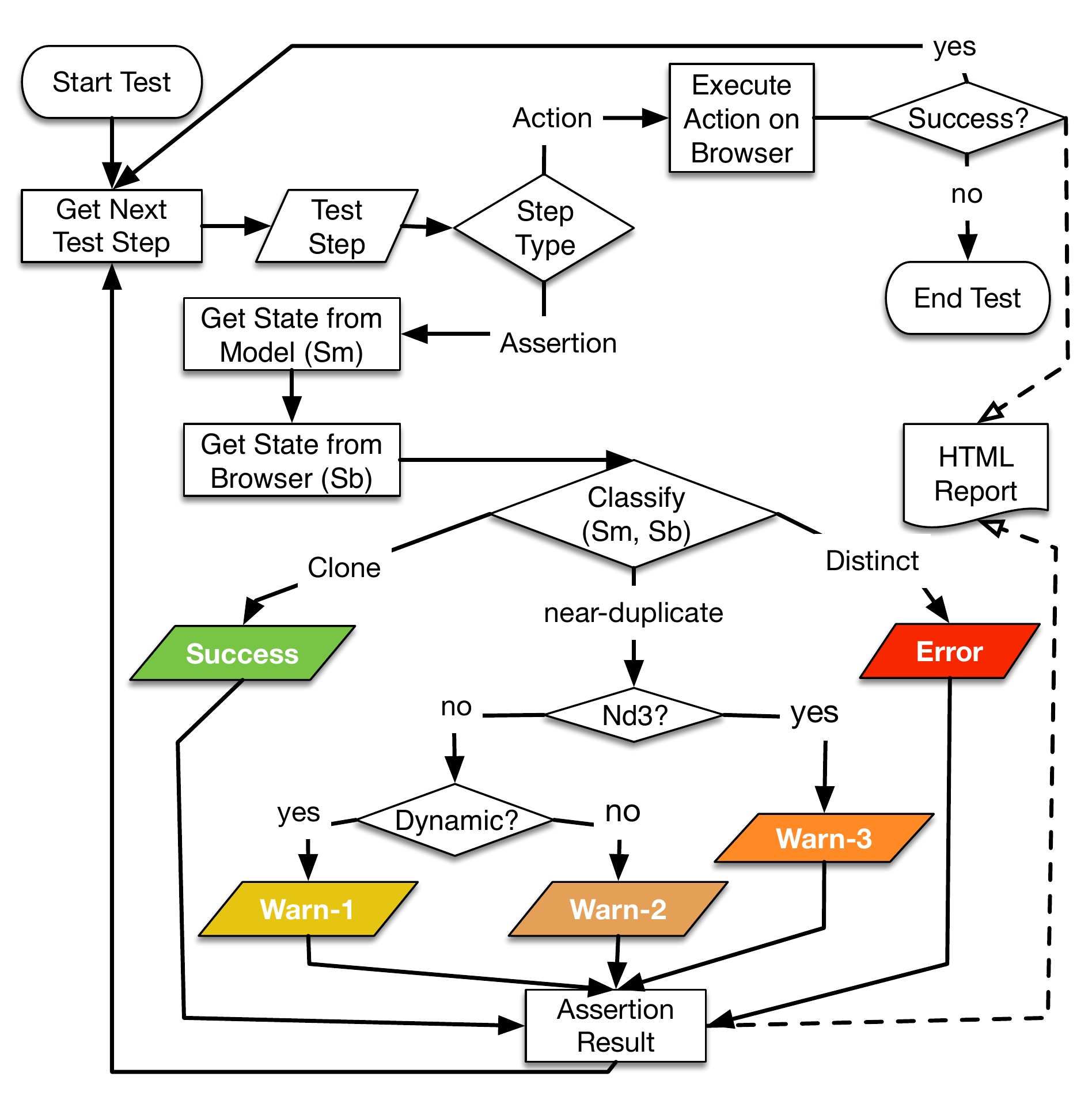}
\caption{\change Test execution flowchart. The function $Classify$ is defined in~\autoref{algo:algo_fragComp} \stopchange{}}
\label{fig:testExecution}

\end{figure}
\change
\subsubsection{Test execution} \label{sec:testExecution}
~\autoref{fig:testExecution} shows the flowchart of our test execution. For each action step, that exercises actionable~(\actionable{}) in the transition~(\transition{}), the test execution continues if successful or stops otherwise. For each assertion step, we invoke our fragment-based comparison~($Classify$ in~\autoref{algo:algo_fragComp}) between the browser state~(\StateVertex{b}) and the model state~(\StateVertex{m}) specified for the particular test step. 
Based on the output of $Classify$ and using \emph{memoization}, \toolname can create three levels of warnings in addition to success and error for each assertion.  Warn-3 has the highest severity while Warn-1 is the lowest.
To the best of our knowledge no prior technique exists to generate such warnings for UI test cases. Existing techniques are only capable of creating a binary decision for a given test oracle and generate either an error or a success status. 

\header{Using data-fluid fragments} We lower the severity of a warning from Warn-2 to Warn-1 when we detect that the changes observed belong to a fragment that is data-fluid. For example, a web element displaying current time on a web page changes in every concrete instance of the web page and should be given a lower severity warning compared to a change to the title of a web page, which has never been found to change during model inference. It is important to recognize that both these changes would just be textual changes in the DOM for any state comparison technique. 

\header{Configuration} In our experiments, we consider any warning other than Warn-1 to be an assertion failure. However, we provide the capability to configure the assertion  failures based on warning levels. Our empirical evaluation assesses how the usage of memoization and identification of data-fluid fragments can help make our assertions robust to near-duplicates while retaining the ability to detect bugs.

 \header{Report}
  \toolname also generates an HTML test report that shows the test execution results  for easier manual analysis. The report provides a visualization of changes and their severity. Our replication package contains a test report for every test execution in the evaluation. 
  \stopchange{}

\subsection{Implementation}
\change
For page fragmentation, existing implementations of VIPS
 rely on web rendering frameworks that are no longer maintained and render modern pages incorrectly. 
 Therefore, we ported one implementation~\cite{vips_impl} to WebDriver API in Java, which is also used by \crawljax so that pages rendered on modern browsers can be fragmented.
\stopchange

We use our  ported version of the VIPS~\cite{vips} algorithm for decomposing a web page into fragments, and compare the structural and visual aspects of the fragments using  APTED~\cite{apted} and Histogram~\cite{Swain} respectively.

We modified the latest version of \crawljax~\cite{mesbah:tweb12} to employ our state abstraction for state equivalence and use our exploration strategy.
  Finally, our test generator is implemented as a \crawljax plugin and generates JUnit test cases using the WebDriver API. Our tool, \toolname, is publicly available~\cite{toolLink}.


\section{Evaluation}

Our empirical evaluation aims to assess \toolname through \change(1) its ability to detect near-duplicates,\stopchange{} (2) the adequacy of its inferred web app models, and (3) the suitability of its generated tests for regression testing.
We do so by answering the following research questions.

\begin{description}[leftmargin=1em]
%
\item \change
\textbf{\q{1}:}
\textit{How effective is \toolname in distinguishing near-duplicates from distinct states  compared to current whole-page techniques?}
\stopchange

\item \textbf{\q{2}:}
\textit{How do the models generated by \toolname compare to the models generated by current techniques?}

\item \textbf{\q{3}:}
\textit{Are the tests generated by \toolname suitable for regression testing?}
\begin{itemize}[leftmargin=1.5em]
\item \textbf{\q{3a}:}
\textit{How do the generated test cases perform in regression testing scenarios?}

\item \textbf{\q{3b}:}
\textit{How effective and tolerant are the generated test oracles?}
\end{itemize}
\end{description}

\change
In \q{1}, we assess the effectiveness of competing techniques in detecting near-duplicates through state-pair classification.\stopchange{}
In \q{2}, we measure the model quality which directly influences the completeness and redundancies in the generated tests. 
In \q{3}, we 
  assess the suitability of generated test suites in regression test scenarios by measuring their reliability in addition to the effectiveness in detecting app changes. 
For \q{3a}, we execute the generated tests on the same web app version but different platform/browser versions to analyze their usability in regression test scenarios,
%
and then evaluate their effectiveness in detecting app changes for a different version of the web app.
 Finally, for \q{3b}, we perform mutation analysis of recorded web states to assess the effectiveness and robustness of generated test oracles.

%
%
\begin{table}[]
\renewcommand{\arraystretch}{1.0}
\caption{Experimental subjects}
\centering
\footnotesize
\label{table:table_subjects}
\resizebox{0.95\columnwidth}{!}{

\begin{tabular}{lrrlr}
\toprule
\textbf{App} & \textit{\textbf{$v_0$}} & \textit{\textbf{$v_1$}} & \textbf{Framework} & \textbf{LOC} \\
\midrule
\textit{\textbf{addressbook}}~\cite{addressbook} & \textit{8.2.5} & \textit{8.2.5.1} & \emph{PHP, JavaScript} & 32K \\
\textit{\textbf{petclinic}}~\cite{petclinic} & \textit{6010d5} & \textit{4aa89ae} & \emph{Java, Spring MVC} & 6K \\
\textit{\textbf{ppma}}~\cite{ppma} & \textit{0.6.0} & \textit{0.5.2} & \emph{Yii, JavaScript} & 556K \\
\textit{\textbf{dimeshift}}~\cite{dimeshift} & \textit{261166d} & \textit{44089e} & \emph{Backbone.js, JQuery} & 10K \\
\textit{\textbf{claroline}}~\cite{claroline} & \textit{1.11.10} & \textit{1.11.9} & \emph{PHP, JavaScript} & 340K \\
\textit{\textbf{phoenix-trello}}~\cite{phoenix} & \textit{60c874d} & \textit{c1cdf30} & \emph{Phoenix, Elixir, ReactJS} & 5K \\
\textit{\textbf{pagekit}}~\cite{pagekit} & \textit{1.0.16} & \textit{1.0.14} & \emph{Symfony, Vue.js} & 275K \\
\textit{\textbf{mantisbt}}~\cite{mantisbt} & \textit{1.1.8} & \textit{1.2.1} & \emph{PHP, JavaScript} & 120K\\
\bottomrule
\end{tabular}
}
\end{table}

\subsection{Subject Systems}\label{sec:subject-set}

To address our research questions,
we needed to manually analyze captured application states which requires a certain level of control over app behaviour and ability to replicate app states under similar experimental conditions. 
\change
In addition, for \q{1} and \q{2}, we require a manually labelled ground-truth for state-pair classification and unique states in a web app respectively. 
\stopchange{}
To this end, we selected eight open-source web apps with an available ground-truth~\cite{ndStudyDataset} and used in prior  web testing research~\cite{2016-Stocco-ICWE,2017-Stocco-SQJ,2019-Biagiola-FSE-Diversity,2019-Biagiola-FSE-Dependecies, ndStudy2020}. 
\change 
Our eight subjects shown in~\autoref{table:table_subjects} 
cover a diverse set of several popular back-end and front-end web app frameworks such as \emph{Symfony, Yii, Spring MVC, Backbone.js, Vue.js, Phoenix/React, JQuery, Bootstrap} and \emph{AngularJS}. Some of our subjects such as \emph{MantisBT, Claroline, PageKit} are quite complex and immensely popular with a sizeable active user base. For example, \emph{Claroline} is an award winning learning management system (LMS) used in more than 100 countries and is available in 35 languages; \emph{MantisBT} is one of the most popular open source issue tracking systems in active use.
\stopchange{}
\subsection{Competing techniques}\label{sec:competingTechniques}

Based on the results of our recent empirical study~\cite{ndStudy2020}, we choose two of the best whole-page techniques, (1) 
RTED~\cite{Pawlik:2015:ECT:2751312.2699485}, a DOM tree differencing technique which inferred the best models on an average,
and, (2) 
Histogram~\cite{Swain}, which was the best performing visual algorithm outperforming RTED on several subjects. 
From here on, we refer to RTED as \emph{Structural} and Histogram as \emph{Visual} when presenting our evaluation.

\change
\section{State-pair Classification (\q{1})}
To address \q{1}, we compare \toolname to whole-page techniques~(\autoref{sec:competingTechniques}) in terms of their ability to classify a given state-pair as either clone, near-duplicate, or distinct.

\begin{table}[t]
\caption{\change Manually classified state-pair dataset \stopchange{}}
\label{table:table_subjectSet}
\centering
\renewcommand{\arraystretch}{1.05}
\footnotesize
\begin{tabular}{lrrrr@{\hskip0.2cm}rrr@{\hskip0.2cm}r}
\toprule

&  \multirow{2}{*}{\rot{\textbf{Pairs}} }& \multirow{2}{*}{\rot{\textbf{Clones}}} & \multicolumn{3}{r}{\textbf{{Near-Duplicates}}} & \multirow{2}{*}{\rot{\textbf{Distinct}}} \\
\cmidrule(r){4-6} 
 
\textbf{Subject}&  & & $Nd_2$ & $Nd_3$ & \textit{Total} &   \\

\specialrule{\lightrulewidth}{8pt}{2pt}

\textit{addressbook} & 8515  & 26   & 52   & 2295  & 2347  & 6142 \\ 
\textit{petclinic}   & 11175 & 2    & 1433 & 180   & 1613  & 9411  \\
\textit{claroline}   & 17766 & 2707 & 69   & 2     & 71    & 14988 \\
\textit{dimeshift}   & 11628 & 375  & 570  & 0     & 570   & 10683 \\
\textit{pagekit}     & 9730  & 0    & 904  & 3044  & 3948  & 5782  \\
\textit{phoenix}     & 11175 & 1    & 25   & 4580  & 4605  & 6569  \\
\textit{ppma}        & 4851  & 64   & 467  & 0     & 467   & 4320  \\
\textit{mantisbt}    & 11325 & 2    & 1117 & 0     & 1117  & 10206 \\
\midrule
Total       & 86165 & 3177 & 4637 & 10101 & 14738 & 68101\\



%

\bottomrule

\end{tabular}

\end{table}

\subsection{Procedure and Metrics}

\header{Dataset} We use an existing dataset (~\autoref{table:table_subjectSet}) of 86,165 state-pairs that are manually labeled as either clone~($Cl$), \ndTwo, \ndThree, or distinct~($Di$) as ground truth. 
It contains 4,637 \ndTwo and 10,101 \ndThree near-duplicates along with 3,177 clones and 68,101 distinct state-pairs belonging to eight subjects used in this work. For the two competing whole-page techniques, \emph{structural} and \emph{visual}, the dataset also includes the computed distance between the two states for each of the 86,165 state-pairs. 

\header{Comparison metrics} 
We use two metrics to compare the competing techniques, 1) the classification $F_1$ score and 2) the number of detected near-duplicates. 
We designed our experiment so that, given a state-pair, each technique classifies it to be either clone~($Cl$), near-duplicate~($Nd$) or distinct~($Di$).
Using the manually classified ground-truth, we then compute
the multi-class classification $F_1$ score which is the average of $F_1$ over all three classes~($Cl, Nd, Di$). 
 
\header{Whole-page technique configuration}
The whole-page techniques by default only output a distance between the two states of a given state-pair. 
In order to compute the  output $class$ for such techniques, we follow previous work~\cite{ndStudy2020} which defined  
a function~$\Gamma$ as :
{\small
$$
   \Gamma(S_1, S_2, \mathcal{W}, t_c, t_n) \left\{
   {\arraycolsep=1.4pt
     \begin{array}{lll}
       Cl&:\mathcal{W}(S_1, S_2) < t_c \\
       &\\
       D&: \mathcal{W}(S_1, S_2) > t_n\\
       &\\
       Nd&:otherwise\
     \end{array}
     }
   \right.
$$
}
$\Gamma$ takes as inputs  a whole-page comparison technique $\mathcal{W}$ which computes the distance between two given states in a state-pair $(S_1, S_2)$ and outputs a $class$ using a pair of thresholds $t_c, t_n$. 
$\Gamma$ classifies a state-pair to be a clone if the computed distance falls below the threshold $t_c$, distinct if it is above $t_n$ or near-duplicate otherwise.
Therefore, the choice of thresholds $t_c, t_n$, can play a huge role in the overall effectiveness of the techniques. 

\header{Threshold determination for whole-page techniques}\label{thresholdDefinition}
To obtain optimal thresholds that maximize the classification scores of the two whole-page techniques, 
using the labelled dataset~(shown in~\autoref{table:table_subjectSet}) as ground-truth, we employ 
 bayesian optimization~\cite{bayesian} to search for thresholds that maximize the multi-class classification $F_1$ score for $\Gamma$.
We ran the optimization technique for 10,000 iterations or trials and retrieved the thresholds that provided best $F_1$ score to be optimal thresholds.  In each trial, the optimizer chooses a pair of thresholds from the sample space of distances possible for the corresponding whole-page technique and computes the $F_1$ score. 

\header{\toolname configuration}
For each state-pair in the dataset, we invoke the $Classify$ function~(\autoref{algo:algo_fragComp}) which outputs one of the four classes~($Di$, $Cl$, \ndTwo, \ndThree). For this experiment, we combine the two outputs \ndTwo and \ndThree into a single class $Nd$ to compare with the output of $\Gamma$. 

\begin{figure}[t]
\centering
%

\begin{subfigure}[b]{0.95\linewidth}
\centering
\includegraphics[trim=0cm 0cm 0cm 0cm, clip=true,  width=0.75\textwidth]{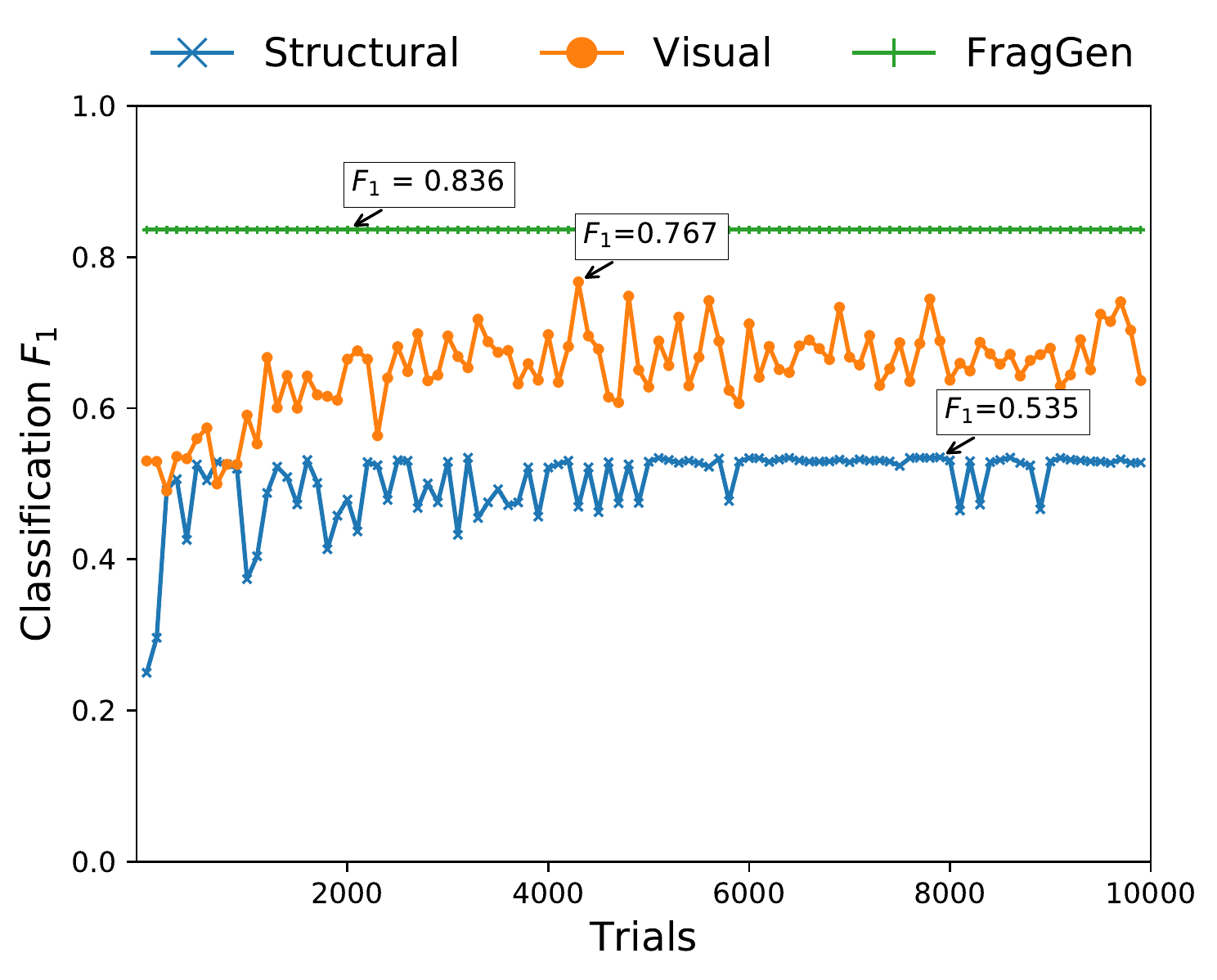}
\caption{\change Finding optimal thresholds using classification $F_1$ \stopchange{}}
\label{fig:fig_thresholdOptimization}
\vspace{1.5em}
\end{subfigure}

\begin{subfigure}[b]{0.95\linewidth}
\centering
\includegraphics[trim=0cm 0cm 0cm 0cm, clip=true, width=0.75\textwidth]{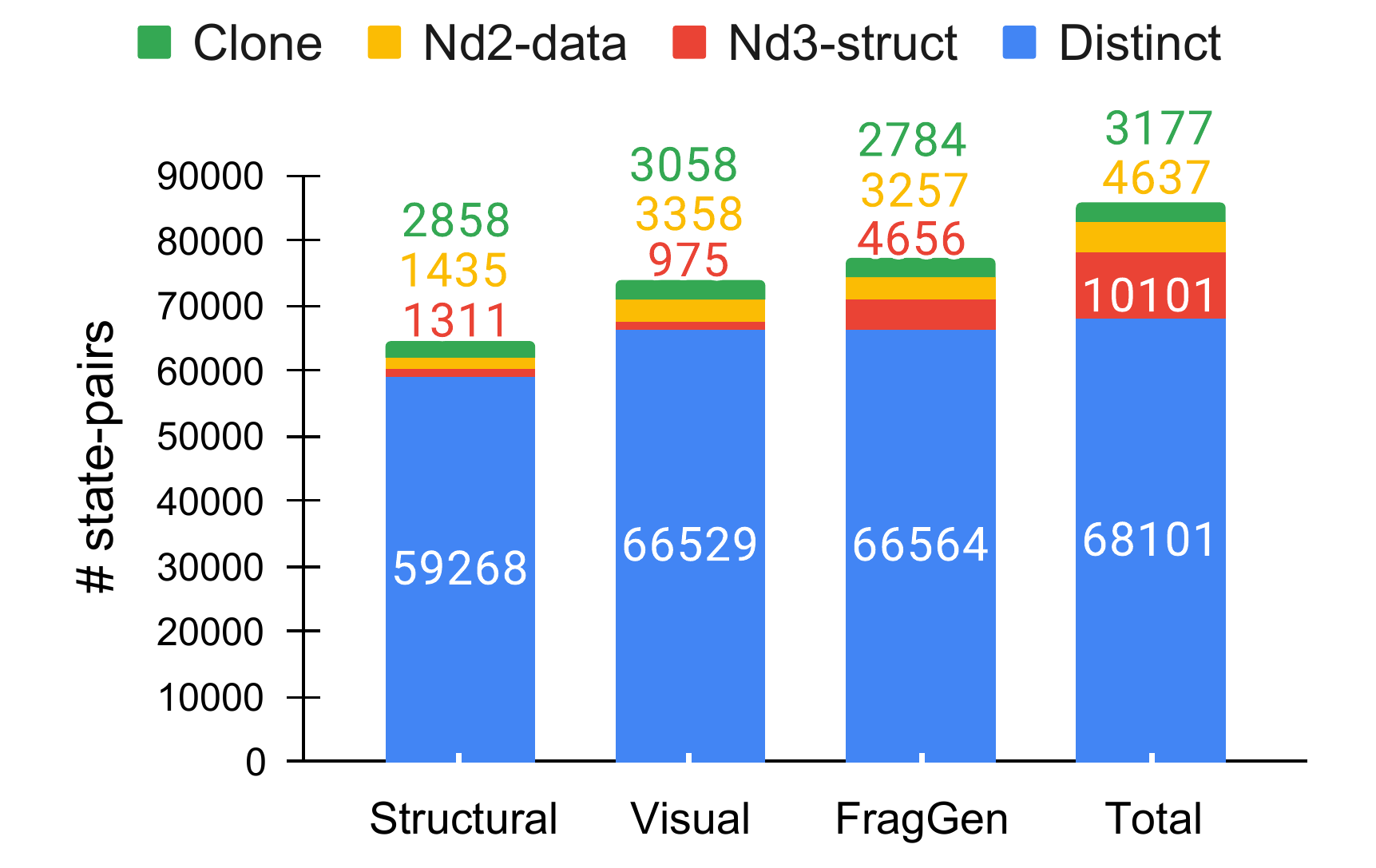}
\caption{\change Correctly classified state-pairs for optimal thresholds \stopchange{}}
\label{fig:fig_classificationBarplot}
\end{subfigure}
\caption{\change State-pair classification results on the dataset \stopchange{}}
\end{figure}

\begin{table*}[!th]
\centering
\caption{$F_1$ of inferred models for 60 minute crawls}
\label{table:table_crawlModelResults}
\footnotesize
\renewcommand{\arraystretch}{1}

\begin{tabular}{lrrrrrrrrr}

\toprule
  & \textit{addressbook} & \textit{claroline} & \textit{dimeshift} & \textit{mantisbt} & \textit{pagekit} & \textit{petclinic} & \textit{phoenix} & \textit{ppma} & \textit{Average} \\

\midrule
\textit{Structural} & 0.44 & 0.83 & 0.26 & 0.42 & 0.40 & 0.79 & 0.38 & 0.23 & 0.47 \\

\textit{Visual}& 0.42 & 0.71 & 0.50 & 0.28 & 0.47 & 0.93 & 0.19 & 0.21 & 0.46 \\

\textit{\toolname} & 0.89 & 0.82 & 0.74 & 0.90 & 0.74 & 1.00 & 0.59 & 0.71 & {0.80}\\

\bottomrule

\end{tabular}
\end{table*}

 \begin{figure*}[!t]
\centering
\includegraphics[trim=3cm 0cm 2cm 0.5cm, clip=true, scale=0.5, height=1.5in, width=5.5in]{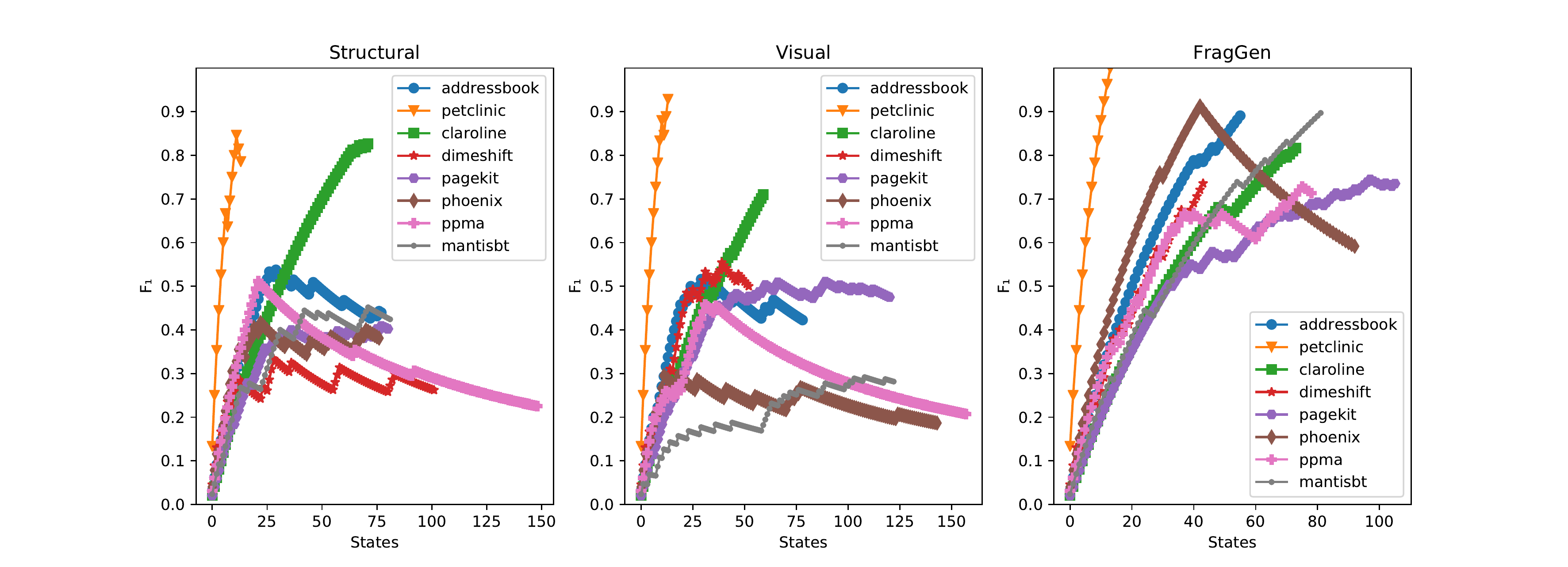}
\caption{$F_1$ as states are detected and added to the model.}
\label{fig:fig_f1}
\end{figure*}

\subsection{Results}

~\autoref{fig:fig_thresholdOptimization} shows the classification $F_1$ being optimized for the structural and visual whole-page techniques by a bayesian optimizer in 10,000 trials.
 To make the data comprehensible, we divided the 10,000 trials into 100 intervals and plotted a single data-point per interval in~\autoref{fig:fig_thresholdOptimization}. 
We chose the maximum $F_1$ found in the corresponding interval of 100 trials as the representative data point of the interval. 

As seen in the~\autoref{fig:fig_thresholdOptimization}, \toolname with an $F_1$ score of 0.836 performs better than the two whole-page techniques. \toolname's $F_1$ score is  55\% better than structural technique and 9\% better than visual technique, which have scores of 0.535 and 0.767 respectively. 


~\autoref{fig:fig_classificationBarplot} shows the correctly classified state-pairs for each of the four state-pair classes in the dataset for the optimal thresholds.
 \toolname correctly classifies the highest number of \ndThree as well as distinct~($Di$) state-pairs, while the visual technique is the best at classifying clones~($Cl$) and \ndTwo state-pairs. Structural is the worst performer overall with a slightly better \ndThree detection compared to visual.  Overall, \toolname correctly classifies 77,261 state-pairs while the visual and structural techniques correctly classify 73,920 and 64,872 state-pairs respectively.

Out of  the 14,738 near-duplicates in the dataset, \toolname detects 7,913 which is 188\% and 82\% better than the structural and visual techniques that detect  2,746 and 4,333 near-duplicates respectively. 
The result is significant because the whole-page techniques performed worse despite having an obvious advantage of classifying the same dataset that is being used for optimization. 
As the previous study~\cite{ndStudy2020} shows, given a set of state-pairs from an unseen web app, the whole-page techniques are unlikely to perform at a similar level using the same thresholds.  

Indeed, the study delves deeper into the threshold selection for generating optimal models; it concludes that 1) the thresholds should be chosen specific to each web app and 2) a ground-truth should be created for any new web app in order to obtain optimal thresholds.
Even after such an optimization per web app, as we illustrate with an example in~\autoref{sec:sec_challenges}, reliance on thresholds creates an inherent limitation for whole-page techniques in separating the \ndThree near-duplicates from distinct state-pairs. 

In \q{2}, we investigate the quality of models inferred by each of the competing techniques configured with the optimal thresholds. 
We obtained optimized thresholds that generate the best model for each of the subjects from the empirical study~\cite{ndStudy2020}.





\stopchange{}

\begin{table}[t]
\centering
\caption{Comparison of inferred models~(\emph{\small eight subjects})}
\label{table:table_nearDupStats}
\footnotesize
\renewcommand{\arraystretch}{1.05}
\resizebox{0.99\columnwidth}{!}{

\begin{tabular}{lrrrrrrrr}
\toprule
\multirow{3}{*}{\textbf{SAF}} & \multicolumn{3}{c}{\textbf{Model Quality}}&\multicolumn{4}{c}{\textbf{Labelled Web States}}& \multirow{3}{*}{\rot{\textbf{Termination}}} \\ 
\cmidrule(l{1em} r{1em}){5-8} \cmidrule{2-4} 

 &\multirow{2}{*}{$Pr$} & \multirow{2}{*}{$Re$} & \multirow{2}{*}{$F_1$} & \multirow{2}{*}{\textbf{Unique}} & \multicolumn{3}{c}{\textbf{Near-Duplicates}} &\\
 
\cmidrule(l{1em} r{1em}){6-8}
 & &  &  &  & $Nd_2$ & $Nd_3$ & \textit{All}   \vspace{1em}\\
 \midrule
Structural & {0.46} & 0.51 & 0.47 & 249 & 44 & 282 & 326 & 2 \\
Visual &  0.45 & {0.52} & 0.46 & 244 & 325 & 116 & {441} & 1 \\
\toolname & {0.79} & {0.83} & 0.80 & {411} & 90 & 29 & 119 & 4 \\
\bottomrule
\end{tabular}
}
\end{table}




\section{Model Inference Comparison (\q{2})} \label{sec:sec_rq1}

\subsection{Procedure and Metrics}
\header{Model quality}
We measure the quality of an inferred model in terms of its coverage of the app state-space, recall ($Re$), and amount of duplication, precision ($Pr$), 
 by manually analyzing it with reference to a ground truth model for the web app using a methodology established in prior research~\cite{ndStudy2020}.
The ground truth models for our subject apps taken from a published dataset~\cite{ndStudyDataset} represent the functionality of a given web app using a minimal set of states and transitions. 
\change A ground truth model is a set of unique states. To analyze a given inferred model, each state in the inferred model is manually \emph{mapped} to one of the ground truth states. A state~(\StateVertex{m}) in the inferred model is mapped to a  state~(\StateVertex{g}) in the ground truth if the manual classification of the state-pair~(\StateVertex{m}, \StateVertex{g}) is either clone, \ndTwo or \ndThree. Every  ground truth state that is mapped to at-least one state in the inferred model is \emph{covered} by the inferred model. We then compute $Pr$ of the inferred model as the ratio of \emph{covered} ground truth states to the total states in the inferred model, and $Re$ as the ratio of \emph{covered} ground truth states to the total number of ground truth states.\stopchange{}

\header{Experiment set-up}
For each of our subject apps, we generate models using each technique by setting a maximum exploration time of 1 hour to be the \emph{stopping criteria}. We use \emph{Google~Chrome}~\emph{(v82.0)} browser, and reset the subject app after every crawl to remove any back-end changes done by previous run. For a fair comparison, we configure  each technique to use exactly the same crawl rules~(e.g., form fill data). \change A technique can also \emph{terminate} exploration before the stopping criterion is invoked by the crawler. \autoref{table:table_nearDupStats} shows the number of times a technique terminated on its own.\stopchange{}




\subsection{Results}
 \autoref{table:table_nearDupStats} shows the overall statistics for all the eight subject apps for \toolname compared to structural and visual techniques. For recall, on average, \toolname covered 83\% of the state space, which is 60\% higher than visual, the next best technique. 
For precision, at 79\%, \toolname produced models with 71\% higher precision than structural, which itself performed slightly better than visual.
 In total, \toolname added only 119 near-duplicates in all models, compared to the 326 by structural and 441 by visual, which are nearly 3 and 4 times higher, respectively. \toolname also discovered \change411\stopchange{} unique states overall, while the existing whole-page techniques detected 37\% less app states in aggregate terms.

 Overall, the F1 measure of \toolname is 0.80, while that of structural and visual whole-page techniques are 0.47 and 0.46, respectively.  \autoref{table:table_crawlModelResults} shows the details for all the subject systems.  \toolname consistently produced models with better $F_1$ except for Claroline where structural technique's model is marginally better. 

When $F_1$ of the models is plotted against the states being added to the model, as shown in~\autoref{fig:fig_f1}, it can be seen that the $F_1$ score for existing techniques does not improve after an initial exploration period.
As a result, the final models generated by existing techniques can sometimes deteriorate over time as more and more near-duplicates are added.
\toolname, however, keeps improving the model quality when given more time as it diversifies the exploration to discover unseen states while avoiding the addition of near-duplicates to the model. 

\begin{figure}[]
\centering
\includegraphics[trim=0.5cm 0.5cm 0.5cm 0.5cm, clip=true, width=0.8\columnwidth]{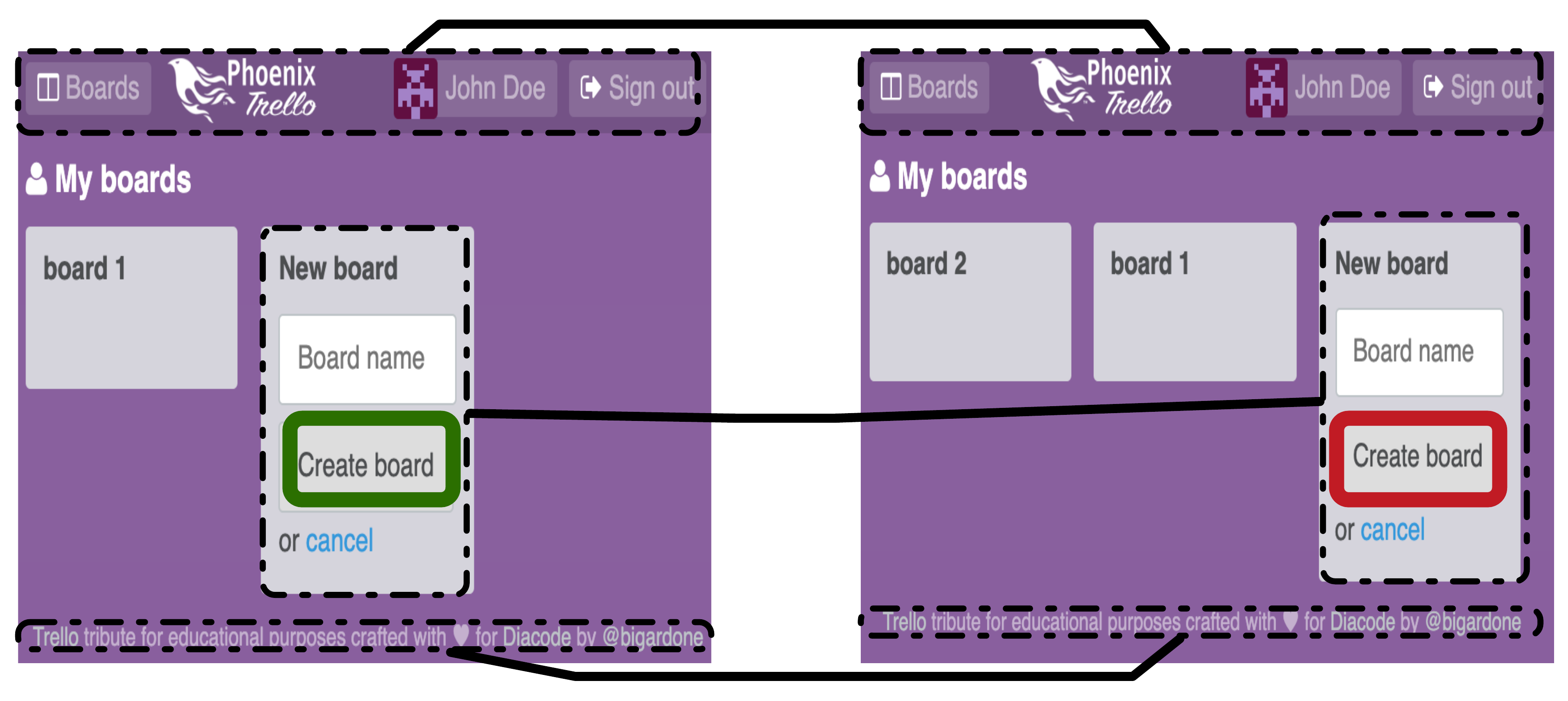}
\caption{Detected near-duplicates in phoenix}
\label{fig:nd_detected} 
\end{figure}

 One of the main reasons for this trend is that existing techniques exercise similar actions repeatedly, and in dynamic web apps, this often results in the creation of near-duplicates and infinite loops, as mentioned in \autoref{sec:sec_challenges}.
Consider a real example from our subject Phoenix, where the action ``create board'' is available in  two model states as shown in~\autoref{fig:nd_detected}. In both states, the action is functionally similar as it creates another board and therefore, need not be exercised more than once. 
 However, existing techniques, which rely on whole-page comparison, cannot infer such similarities 
 and keep creating boards and \ndThree near-duplicate states as they repeatedly exercise the ``create board'' action.  
 
 On the other hand, \toolname is able to identify similar actionables through fine-grained fragment-based analysis to avoid creating \ndThree states and successfully diversify the exploration. 
 \change Indeed, as the~\autoref{table:table_nearDupStats} shows, \toolname added just 29 \ndThree states overall while \emph{structural} and \emph{visual} techniques added 282 and 116 respectively. 
 
 The trend is prominently noticeable in~\autoref{fig:fig_f1}, where \toolname is able to improve the model continuously by avoiding addition of \ndThree near-duplicates to the model and seeking out unique actionables to exercise.\stopchange{}
As a result, \toolname also terminates exploration for 4 out of 8 subjects apps as shown in ~\autoref{table:table_nearDupStats}, whereas structural and visual techniques terminate only 2 and 1 times respectively.

\section{Regression Testing Suitability~(\q{3})} \label{sec:sec_rq2}

As part of assessing the suitability of generated tests in regression scenarios, in \q{3a}, we execute generated tests on different browser/platform and app versions, in order to evaluate the model inaccuracies and fragility of test cases in addition to the effectiveness of techniques in detecting application changes. 
In \q{3b}, we further evaluate the robustness and effectiveness of test oracles by simulating evolution through mutation of 
recorded web states.

\subsection{Test Breakages~(\q{3a})} \label{sec:sec_rq2a}

\subsubsection{Procedure and Metrics}
\change
 The goal of our evaluation in RQ3a is to assess how \toolname compares against existing techniques that generate regression tests automatically. There are not many techniques available that generate regression tests for web applications currently. 
 We compare tests generated by \toolname against tests generated by \crawljax, which also employs exploration paths to generate tests from the crawl model.
 \stopchange{}
 While \toolname uses its fine-grained fragment analysis to generate test oracles, for the whole-page techniques,
  \crawljax is configured to  generate test oracles that use the same whole-page comparison as the one used for model inference, namely, structural (RTED) and visual (Histogram). 

\autoref{table:table_regressionExpSetup} shows the three regression test scenarios used in our evaluation, where in $\epsilon_1$ and $\epsilon_2$, we execute tests on the same app version but vary browser/platforms from the crawl, allowing us to evaluate the validity and robustness of the generated test suite. 
We then execute tests~($\epsilon_3$) on a different version of the web app to determine the effectiveness of the test suites in detecting real application changes.  

\begin{table}[]
\renewcommand{\arraystretch}{0.9}

\caption{Regression test execution set-up}
\footnotesize
\centering
\label{table:table_regressionExpSetup}
\begin{tabular}{llll}
\toprule
Execution & app\_version & Platform & Browser  \\
\midrule
Crawl & $v_0$ & MacOS-14~\cite{mojave} & chrome-82 \\
TestSuite $\epsilon_1$ & $v_0$ & MacOS-14 &  chrome-83\\
TestSuite $\epsilon_2$ & $v_0$ & RHEL-7~\cite{rhel7} &  chrome-84\\
TestSuite $\epsilon_3$& $v_1$ & RHEL-7 & chrome-84\\
\bottomrule

\end{tabular}
\vspace {-1em}

\end{table}

 \begin{figure}[]
\centering
\includegraphics[trim=0.5cm 0cm 1cm 0cm, clip=true, scale=0.4]{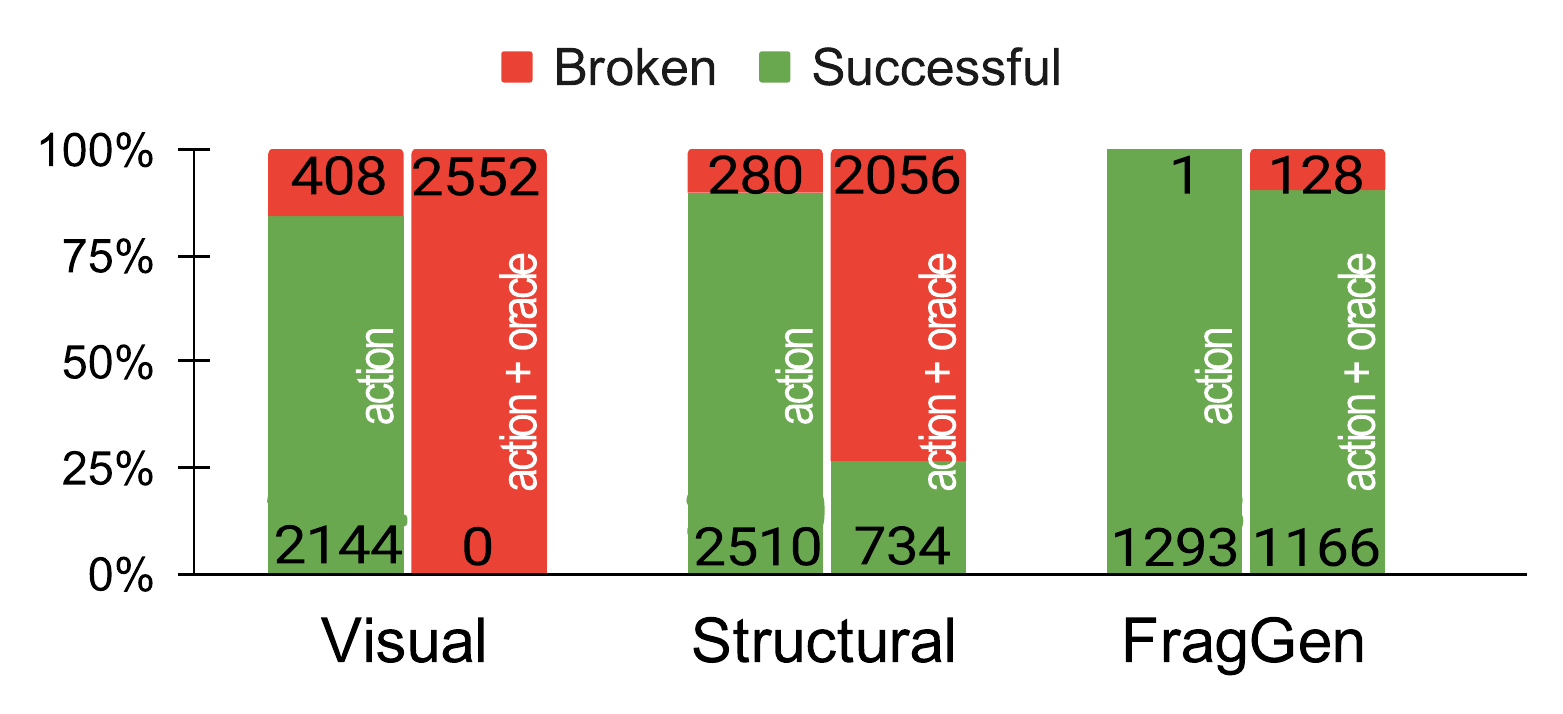}
\caption{Test breakages on the same app version~($\epsilon_1, \epsilon_2$) }
\label{fig:fig_sameVersion_barplot}
\end{figure}

\begin{table*}[]
\centering
\renewcommand{\arraystretch}{1.05}
\footnotesize
\caption{Regression test run results}
\label{table:table_regression}
\begin{tabular}{l@{\hskip 0.7cm}rrrr@{\hskip 0.7cm}rrrr@{\hskip 0.9cm}rrr@{\hskip 1cm}rrrr}

\toprule

\multicolumn{1}{c}{\textit{}} & \multicolumn{11}{c}{\textbf{Whole-page techniques}} & \multicolumn{4}{c}{\multirow{2}{*}{\textbf{\toolname}}} \\

\cmidrule(l{0em} r{3em}){2-12}
\multicolumn{1}{c}{\textit{}} & \multicolumn{4}{l}{\textbf{~~~~~~~~~~~~Visual}} & \multicolumn{4}{l}{\textbf{~~~~~~~~~Structural}} & \multicolumn{3}{l}{\textbf{~~~~~~~Average}} & \multicolumn{4}{c}{} \\

\cmidrule(l{0em} r{3em}){2-5} \cmidrule(l{0em} r{3em}){6-9} \cmidrule(l{0em} r{3em}){10-12}\cmidrule(l{1em} r{1em}){13-16}
\textit{\textbf{Test Execution}} & \multicolumn{1}{c}{\textit{\textbf{$\epsilon_1$}}} & \multicolumn{1}{c}{\textit{\textbf{$\epsilon_2$}}} & \multicolumn{1}{c}{\textit{\textbf{$\epsilon_3$}}} & \multicolumn{1}{l}{\textit{\textbf{All}}}& \multicolumn{1}{c}{\textit{\textbf{$\epsilon_1$}}} & \multicolumn{1}{c}{\textit{\textbf{$\epsilon_2$}}} & \multicolumn{1}{c}{\textit{\textbf{$\epsilon_3$}}} & \multicolumn{1}{l}{\textit{\textbf{All}}} & \multicolumn{1}{c}{\textit{\textbf{$v_0$}}} & \multicolumn{1}{c}{\textit{\textbf{$v_1$}}} & \multicolumn{1}{l}{\textit{\textbf{All}}} & \multicolumn{1}{c}{\textit{\textbf{$\epsilon_1$}}} & \multicolumn{1}{c}{\textit{\textbf{$\epsilon_2$}}} & \multicolumn{1}{c}{\textit{\textbf{$\epsilon_3$}}} & \multicolumn{1}{l}{\textit{\textbf{All}}} \\

\midrule
\textit{\textbf{Action Success \%}} & 83.6 & 83.2 & 63.0 & 76.6 & 89.1 & 88.8 & 69.0 & 82.3 & {86.2} & 66.0 & 79.5 & \textit{{99.5}} & \textit{{99.9}} & 93.3 & {97.6} \\
\textit{\textbf{Oracle Success \%}} & \textit{{0.0}} & \textit{{0.0}} & \textit{{0.0}} & \textit{{0.0}} & 53.1 & 51.3 & 22.8 & \textit{{42.4}} & {26.1} & 11.4 & 21.0 & \textit{{98.6}} & \textit{{97.5}} & 64.8 & {87.0} \\
\bottomrule
\end{tabular}
\end{table*}

\subsubsection{\q{3a} Results}
As seen in ~\autoref{table:table_regression}, 
whole-page techniques generated test actions that succeeded only 86.2\% on an average.
A potential cause for failure of remaining nearly 14\% actions on the same version could be limitations in handling the near-duplicates for existing techniques. 
As ~\autoref{fig:fig_sameVersion_barplot} show, breakage of these test actions  
 resulted in breakage of 16\% and 10\% of tests in visual and structural test suites respectively without even considering the test oracle fragility.  
When test oracles are considered in declaring test breakages,  all visual test oracles fail during test executions causing 100\% test breakage, while 52\% structural oracles fail breaking nearly 74\% of tests in the same app version.

In contrast, \toolname can execute test actions 
on the same version of the web app with nearly 100\% success rate. As~\autoref{table:table_regression} shows, in $\epsilon_1, \epsilon_2$, \toolname's test oracles also have a 98\% success rate on average, showing greater adaptability to changing execution environments compared to the whole-page techniques. 

\begin{figure}[h!]
\begin{subfigure}[b]{0.5\textwidth}
\centering
\includegraphics[trim=0cm 0cm 0cm 0.cm, clip=true, scale=0.32]{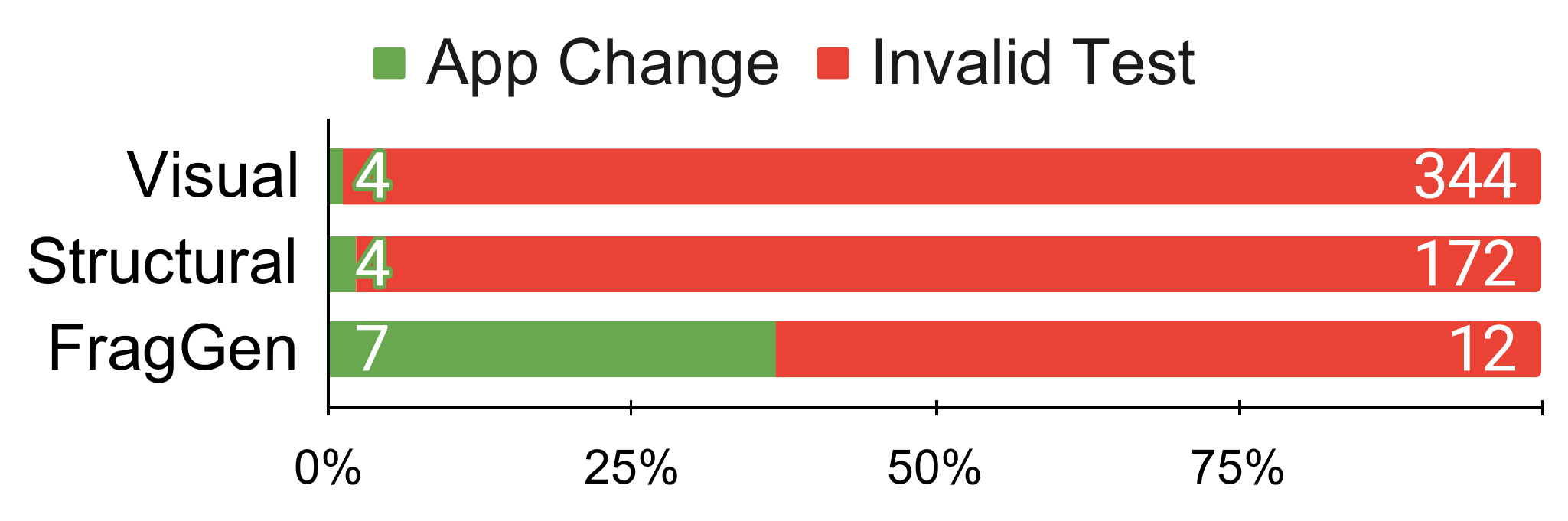}
\caption{Test action failures~($\epsilon_3$)}
\vspace{0.4em}
\label{fig:fig_appVersion_barplot}
\end{subfigure}

\begin{subfigure}[b]{0.5\textwidth}
\centering
\includegraphics[trim=0cm 0cm 0cm 0.cm, clip=true, scale=0.32]{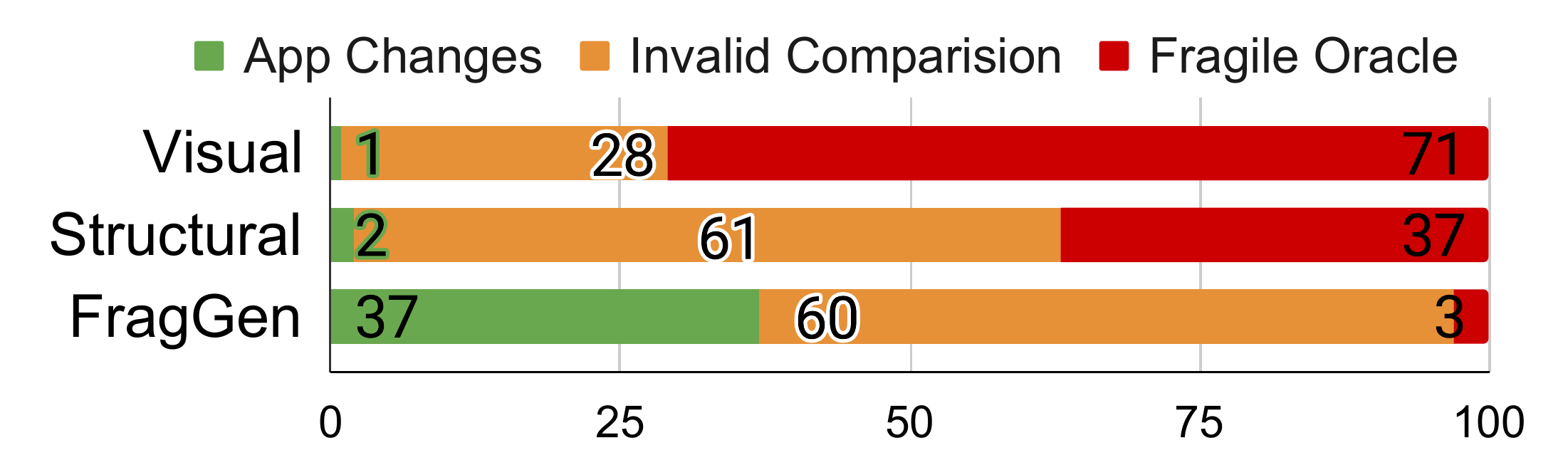}
\caption{Random test oracle failures~($\epsilon_3$)}
\label{fig:fig_testOracles_barplot}
\end{subfigure}
\caption{Manual analysis of failed test actions and oracles on a different version of the web apps}
\end{figure}

Next, we evaluate test suite effectiveness in detecting app changes on a different version of web app using the test execution~($\epsilon_3$).
We follow the standard web testing practice where tests created for an existing version~($v_0$) are executed on a new version~($v_1$) of web app in order to detect faults through manual analysis of test failures.
We mitigate the potential bias in manual analysis of test failures by labelling
 \emph{all} observed application changes to be faults. 

When we manually analyzed all test failures in $\epsilon_3$ as shown in~\autoref{fig:fig_appVersion_barplot}, we found that 
 both structural and visual test suites detected 4 app changes each and produced 172 and 344 false positives respectively.
In contrast, \toolname was able to detect 7 app changes with only 10 false positives, significantly reducing the manual effort required to identify app changes and maintain test suites.



\begin{table*}[t]
\centering
\setlength{\tabcolsep}{4pt}
\renewcommand{\arraystretch}{1}
\caption{Mutation operators for DOM nodes}
\label{table:table_mutationOperators}
\footnotesize
\resizebox{0.9\linewidth}{!}{
\begin{tabular}{llll}
\toprule

\textbf{Type} & \textbf{Tags or Attributes}& \textbf{Description} & \textbf{Example} \emph{(original : mutant)}\\ 

\midrule

Attribute  & \{id, class, title\} & Modifies mentioned attribute value of any node. & [\textless{}div id="a"\textgreater{}] : [\textless{}div id="aMut"\textgreater{}] \\
Tag  & \{span, h1-h6, p\} & The tag name is changed to a similar tag. & [\textless{}h1\textgreater{}xyz\textless{}/h1\textgreater] : [\textless{}h2\textgreater{}xyz\textless{}/h2\textgreater{}]  \\
Subtree  & \{div, table, tr, td, ul, li, p\} &  Deletes all children of selected container node.& [\textless{}tr\textgreater{}\textless{}td\textgreater{}xyz\textless{}/td\textgreater{}\textless{}/tr\textgreater] : [\textless{}tr\textgreater{}\textless{}/tr\textgreater{}] \\
Text  & \{h1-h6, p, b, i, \} & The text content of selected leaf node is changed. &[\textless{}h1\textgreater{}abc\textless{}/h1\textgreater] : [\textless{}h1\textgreater{}abcMut\textless{}h1\textgreater{}]\\

\bottomrule
\end{tabular}
}
\end{table*}

We then manually analyzed 100 randomly selected oracle failures for each technique to label the cause of failure.
In this manual analysis, if the test state matches the expected state, we categorize the failure as either an \emph{app change} if we notice a change in application behaviour or a \emph{fragile oracle} as the failure represents a limitation of state comparison. 
Our analysis depicted in~\autoref{fig:fig_testOracles_barplot}, shows that 
 71 visual and 37 structural oracles failed because of fragility in state comparison, whereas 
 \toolname generated only 3 fragile test oracles. 
Of the 100 examined failures, while the structural and visual oracles only detected 2 and 1 app changes respectively,  \toolname could detect 37 app changes.

When the test state and expected state do not match for the failure being analyzed, we categorize it as an invalid comparison, where typically test execution does not reach the target app state due to failure of earlier test actions or invalid test sequences due to changed app behaviour.
Further root cause analysis of such failures is difficult even with domain knowledge, which makes it challenging to directly evaluate test oracle robustness through regression test executions. 
We mitigate this problem in \q{3b}, where the test oracles are compared on a synthetic dataset generated through mutation of recorded web pages.


These results show that the tests generated by \toolname are significantly more robust than existing techniques in regression testing scenarios involving the same version of the web app, and show greater efficacy in detecting evolutionary app changes with significantly lower manual effort.


\subsection{Effectiveness and Robustness of Test Oracles (\q{3b})}

In \q{3b}, we simulate evolutionary app changes using mutation analysis to evaluate 
the suitability of generated test oracles for regression testing.



\subsubsection{Mutation Analysis Methodology}
We define four mutation operators for DOM nodes as shown in ~\autoref{table:table_mutationOperators} in order to generate mutant web pages. 
Similar mutation operators for GUI artifacts have been proposed in prior web~\cite{mutationAnalysis, mutationSalt, mutation_web_offutt}, Android~\cite{mutation_android_linares} and GUI~\cite{mutation_gui, gui_mutation_memon} mutation analysis research.  
Since \q{3b} only aims to evaluate test oracles, we find this approach of  mutating recorded web pages to be adequate and more flexible, instead of mutating the actual web app source code as prior work does. 

Given a crawl model and a test execution trace, our mutation analysis tool applies a random mutation to one of the model states and compares the mutant to the corresponding states in the test trace using each of the three competing techniques.

We partition the generated mutants into either app changes or 
irrelevant to functionality
 based on their visibility in the GUI. 
We categorize (1) the mutants not visible on the page as equivalent mutants that test oracles should tolerate and ignore, and 
(2) all visible mutations as app changes that should be detected by effective test oracles.
\change As~\autoref{table:table_mutationResults} shows, we use 3,042 Visible mutations to compute effectiveness. Robustness is computed using 6,013 equivalent mutants that are of three kinds : \emph{None} - where no mutation has been applied on the state; \emph{AttributeMutator} - an attribute is mutated; and, \emph{Invisible} - where a hidden web element is mutated. \stopchange{}

 Note that we recognize the possibility of invisible DOM or attribute changes affecting web page functionality.
  However, we expect that such changes manifest in failure of test actions, which is outside the scope of \q{3b}, as our set-up does not execute test actions on these mutated states. 
  \change
   Equivalent mutants of the kind \emph{None} may still have DOM changes 
   that result from a variety of reasons such as autogeneration of pages in back-end.
   We still consider them to be equivalent because the changes do not relate to either 1) change in source code which was unmodified or 2) change in test execution which remained consistent.
   \stopchange

Finally, we consider a test oracle to be suitable for regression testing if it has high robustness as well as effectiveness.

\header{\toolname configuration}
As described in~\autoref{sec:memoization}, during model inference, \toolname identifies data-fluid fragments
 and employs this knowledge to generate test oracles.
In \autoref{table:table_mutationResults}, \emph{No-Mem} refers to test oracles if only fragment-based comparison is used; and \emph{With-Mem} shows the test oracles that employ knowledge of data-fluid fragments. 
\change
As shown in~\autoref{fig:testExecution}, without the knowledge of data-fluid fragments, we cannot lower the severity of warnings for changes that are typical for the web app under test~and therefore may not necessarily be application bugs.
\stopchange{}

\begin{table}[]
\centering
\caption{Effectiveness and Robustness of test oracles}
\label{table:table_mutationResults}
\renewcommand{\arraystretch}{1.0}
\footnotesize
\begin{tabular}{l@{\hskip 0.2cm}rr@{\hskip 0.2cm}rr@{\hskip 0.2cm}r@{\hskip 0.25cm}rr@{\hskip 0.15cm}r@{\hskip 0.15cm}}
\toprule
\multirow{2}{*}{\textbf{Score}} & \multicolumn{2}{l}{~~~Mutation}  && \multirow{2}{*}{\rot{\textbf{Structural}}} & \multirow{2}{*}{\textbf{\rot{Visual}}} & &\multicolumn{2}{l}{~~~\toolname} \\
\cmidrule{2-3}
\cmidrule{8-9}
\multicolumn{1}{c}{} &  \multicolumn{1}{c}{\textit{{\rot{Type}}}} &  \multicolumn{1}{c}{{{{\#}}}}  && \multicolumn{1}{l}{} & \multicolumn{1}{l}{} && \multicolumn{1}{r}{\textit{{\rot{No-Mem}}}} & \multicolumn{1}{r}{\textit{{\rot{With-Mem}}}}\\
\midrule
&&&&&&& \\
& \textit{\textbf{Visible}} & \textit{3042} && 2235 & 2908  && 3001 & 3001 \\
\cmidrule{2-9}
\textbf{Effectiveness} &  & \multicolumn{1}{l}{\textit{}} && \textbf{73.5} & \textbf{95.6}  && \textbf{98.7} & \textit{\textbf{98.7}}\\
\midrule

&&&&&&&\vspace{0.1em} \\
 & \textit{\textbf{None}} & \textit{4620} && 198 & 1964  && 921  & 136\\
 & \textit{\textbf{Attribute}} & \textit{770} && 10 & 585   && 166& 157 \\
 & \textit{\textbf{Invisible}} & \textit{623} && 553 & 577 && 275 & 274  \\
 &  \textit{\textbf{Total}}& \textit{6013} && 761 & 3126  && 1362 & 567 \\
\cmidrule{2-9}
\textbf{Robustness} &  & \multicolumn{1}{l}{\textit{}} && \textbf{87.3} & \textit{\textbf{48.0}} && \textit{\textbf{77.3}}& \textbf{90.6}  \\

\bottomrule

\end{tabular}
\end{table}

\begin{figure}[t]
\centering
\includegraphics[trim=0.5cm 0cm 1cm 1cm, clip=true, width=0.8\columnwidth]{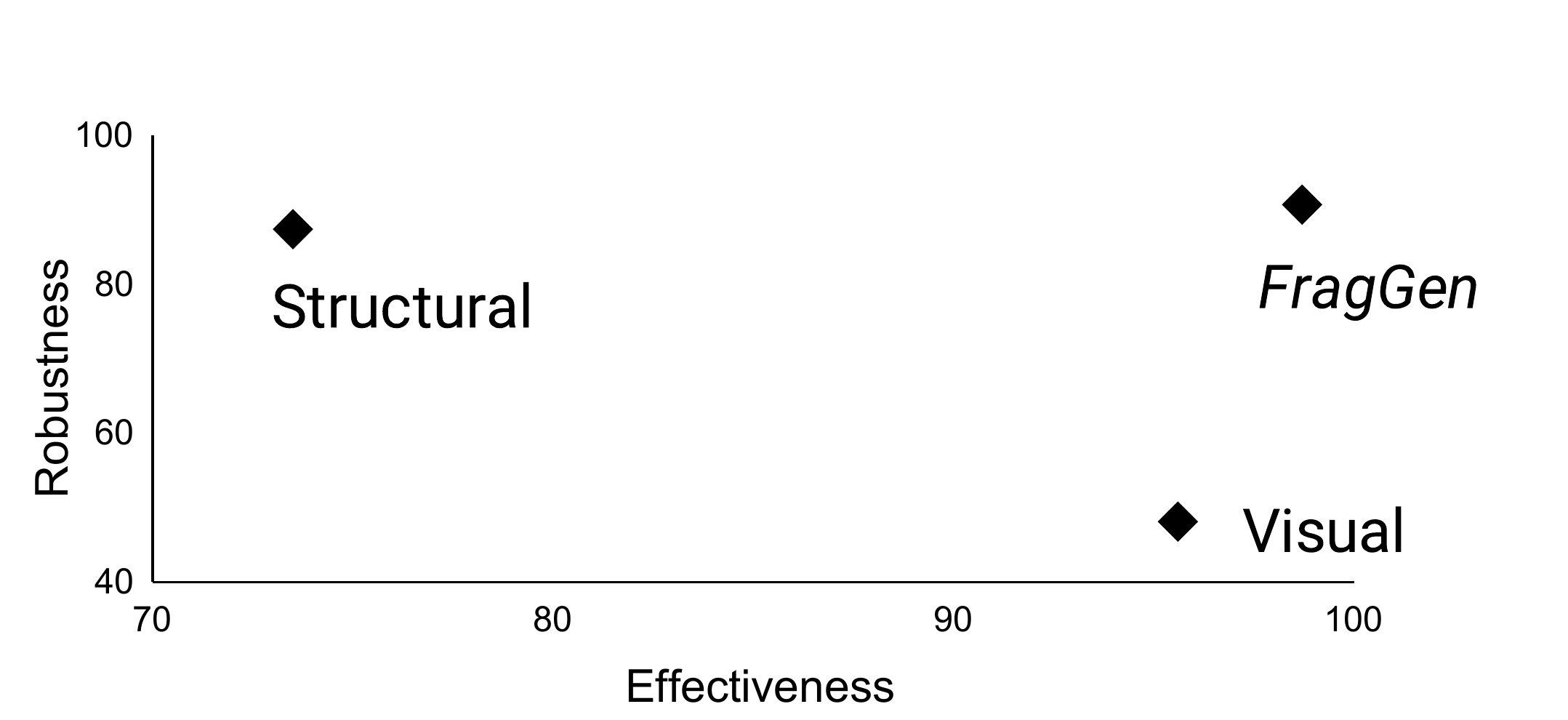}
\caption{Effectiveness vs Robustness of test oracles}
\label{fig:fig_mutationChart} 
\end{figure}

\subsubsection{Results}
 
As~~\autoref{table:table_mutationResults} shows, our mutation analysis experiment resulted in a total of 3,042 randomly generated visible mutations that should be detected by the oracles and 6,013 equivalent mutants that the oracles should be tolerant to. 



Overall, amongst the existing techniques, \emph{visual} test oracles detect 95.6\% of visible mutations while \emph{RTED}, the structural comparison technique which only considers HTML tags for equivalence, is tolerant to 87.3\% of equivalent mutants encountered during the experiment. 
However, visual oracles are fragile, with a robustness score of only 48\%, whereas structural oracles fail to detect 26.5\% of the visible mutants.

In contrast, test oracles generated by \toolname are able to detect 98.7\% of visible mutants in both the configurations.
When \emph{memoization} knowledge is applied in the \emph{With-Mem} configuration, \toolname has the highest robustness of all the competing techniques by detecting 90.6\% of the equivalent mutants. 
In the \emph{No-Mem} configuration, when only fragment-based state comparison is used, 
 the robustness drops to 77.3\%, 
 which is still 61\% more than visual. 
The difference in numbers between the two configurations is because of handling the \emph{None} category, where no mutation has been applied to the recorded web pages. It can be concluded therefore, that \toolname is able to make use of memoization and data-fluid fragments to improve the robustness of test oracles. 

As the \autoref{fig:fig_mutationChart} shows, existing whole-page oracles can either be highly effective or highly tolerant but cannot balance both the aspects, causing either a large number of false positives or false negatives respectively.
 \toolname's test oracles on the other hand, use fine-grained fragment analysis with memoization to 
 outperform existing techniques in both effectiveness and robustness at the same time,
%
making them suitable for regression testing of modern web apps.





\section{Discussion}

\header{Granularity of fragments}
During our experiments, we found that 
determining equivalency of 
certain fragments that are too small such as \Fragment{3}, shown in dotted lines in~\autoref{fig:fig_fragmentation}, 
 using DOM or visual characteristics results in drawing false equivalence between semantically different web pages. 
In order to avoid this, using a predefined threshold, \toolname prunes out the smaller fragments and does not use their equivalence in determining page similarity. 

%

\begin{figure}[]
\centering
\includegraphics[trim=0.5cm 0.5cm 0.5cm 2cm, clip=true, width=0.96\columnwidth, height=1.6in]{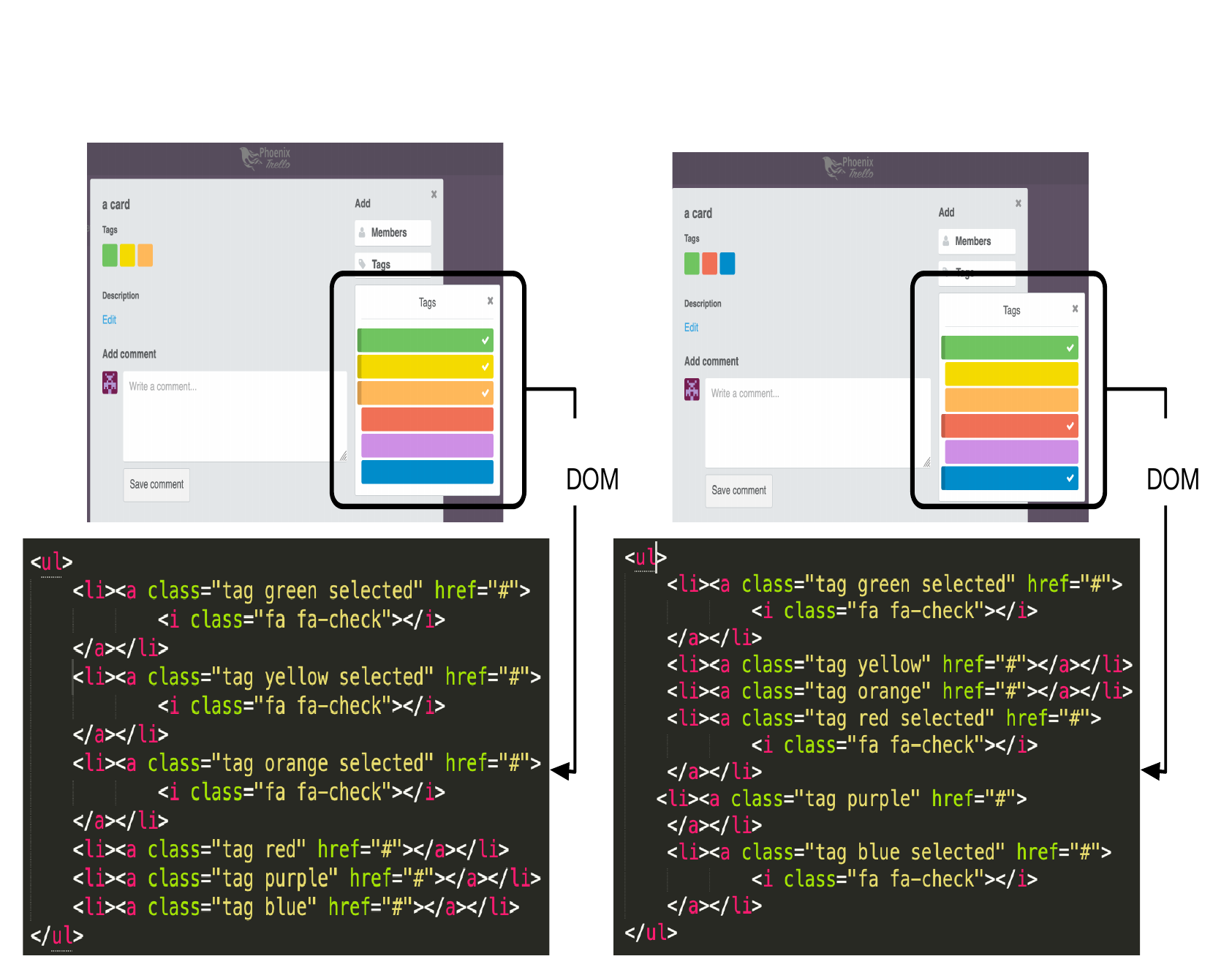}
\caption{Undetected near-duplicate fragments in phoenix}
\label{fig:fig_phoenixND}
\end{figure}

 One such example is shown in \autoref{fig:fig_phoenixND}, where each individual tag can have two states -- selected or not. Therefore each tag can be semantically equivalent to other depending on this state.
  However, as 
  we avoid comparing the smaller fragments of individual tags, we cannot determine equivalency of parent fragments which differ structurally upon selection of different tags.
 The number of possible near-duplicates we cannot detect in such cases can explode in number, 
  and is the reason for the reduced precision of the web app model for \emph{phoenix}.

Even-though smaller fragments with even one web element can be separate functional entities, such fine-grained comparison needs semantic inference beyond simple DOM and visual characteristics, which we leave for future work. 

\header{Test oracle generation}
\toolname currently identifies data-fluid fragments assuming that during model inference, only the events fired by the crawler can cause persistent changes. Such assumption can fail on live web applications where multiple users can concurrently induce back-end changes. However, since our technique is meant for test environments, we consider our assumption reasonable.
\change
\header{Model Inference in other domains}
While our \toolname implementation is specific to web apps, conceptually our approach to model generation based on fragments (instead of whole pages or whole screens) can be applied to other domains such as mobile apps or desktop applications.
\stopchange{}


\header{Threats to validity}
Using a limited number of web apps in a controlled setting in our evaluation poses an external validity threat and further experiments are needed
to fully assess the generalizability of our results; we have chosen
eight subject apps used in previous web testing research, pertaining
to different domains in order to mitigate the threat.
Threats to internal validity come from 
 the manual labelling of web pages and changes, which was unavoidable because no automated method could provide us with the required ground truth. 
 To mitigate this threat, we performed the labelling by following a process established in prior work. 
For reproducibility of our findings, we made our tool publicly available~\cite{toolLink} along with usage instructions and used subject systems.

\section{Related Work}

In web model inference, \crawljax~\cite{mesbah:tweb12} explores dynamic web apps with client-side JavaScript by triggering actions and analyzing the DOM. 
WebMate~\cite{webmate} crawls web apps using a state equivalence based on actionables in web pages.
 \toolname is built upon \crawljax framework by adding a new fragment-based state abstraction and modifying  several core components that drive state exploration.

For regression test suite generation from web app models, Mesbah et al. proposed test generation based on model coverage~\cite{mesbah:tweb12} while Marchetto et al.~\cite{4539539} used coverage criteria based on semantically interacting events. \toolname uses the exploration paths that cover all the events and states in the inferred model for test generation. 

Instead of crawl models, Biagiola et al. use \emph{page objects} derived from crawl models, and employed search-based~\cite{subweb} techniques, diversification~\cite{2019-Biagiola-FSE-Diversity}  of test events to generate test cases.  These page objects used in existing work are derived from crawl models generated by the baseline Crawjax that was manually configured with thresholds by the researchers for their experiments.
 In contrast, we generate test cases directly from crawl models without any manual threshold selection using the paths followed by our exploration strategy during model inference. 
Although the generation of page objects from crawl models is largely automated, some manual effort such as specifying guards to avoid test dependencies is required for optimal results.
 Indeed, the superior crawl models generated by \toolname should improve the test cases generated by the existing techniques based on page objects.



Prior research used histogram~\cite{Choudhary-2012-ICST} to detect cross browser differences  and manually specified  DOM invariants~\cite{Mesbah:2009} to create robust test oracles.
\toolname, however, automatically generates robust fine-grained test oracles at fragment level by leveraging model inference. 

To assess the quality of web test suites, 
prior work mutated source code artifacts such as JSP web pages~\cite{mutation_web_offutt,mutation_web_offutt_eval}, and client-side JavaScript~\cite{mutationSalt,mutationAnalysis}. We assess the quality of test oracles by mutating the recorded web pages. 

Sun et al.~\cite{vipsCrawler} employed page fragment-based exploration for efficient information retrieval while we use fragmentation for model inference and test oracle generation.

\change{}
FeedEx~\cite{amin:issre13} employs a combination of client-side JavaScript code coverage, DOM and Path diversity to prioritize re-exploration of already explored actions. WebExplor~\cite{reinforcementCrawler} by Zheng et al. also proposes a method to prioritize web elements that were already explored by rewarding discovery of new states in previous result of the same action. Both these existing techniques assign the same priority to all unexplored actions. 
 State exploration strategy is a concern for UI testing in general and Degott et al.~ \cite{zellerMobile} also prioritize exploratory actions by first exploring all available actions in a mobile app and analyzing the results of execution. In contrast, FragGen is able to make use of fragment based equivalence of unexplored actions to diversify exploration.

\stopchange{}

 To the best of our knowledge, we are the first to employ page fragmentation to establish state equivalence for web app testing. Our novel state comparison combines both structural and visual analysis of the identified fragments to effectively detect near-duplicates. 
Our tool, \toolname, uses this novel state comparison to
infer precise models, as well as  
   generate effective and reliable regression test suites. 

\section{Conclusions and Future Work}\label{sec:conclusions}
Automated model inference and test generation for complex dynamic modern web apps is a challenging problem because of the presence of near-duplicates that cannot be detected by whole-page analysis employed by existing techniques. We developed a novel technique, \toolname, which uses smaller page fragments to detect near-duplicates and diversify web app exploration to generate precise models while covering a high percentage of application state space. \toolname is also able to generate oracles that are suitable for regression testing as they are highly effective in detecting visible app changes while being tolerant to minor changes unrelated to functionality.

As part of the future work, we plan to improve our state comparison technique through inference of web page semantics when structural and visual characteristics are inadequate.

\bibliographystyle{IEEEtran}
\bibliography{paper}

\begin{IEEEbiography}
    [{\includegraphics[width=1in,height=1.25in,clip,keepaspectratio]{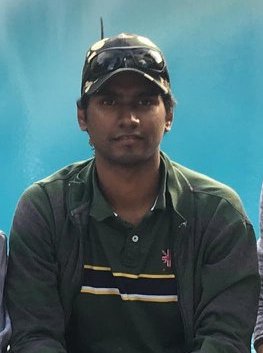}}]{RahulKrishna Yandrapally}
 received his Bachelors and Masters in Computer Science and Engineering from Indian Institute of Technology, Kanpur
(IIT Kanpur) in 2012. 
Thereafter, he joined the Software Engineering group at IBM Research, India as a blue scholar and developed tools for web and mobile testing. 
Currently, he is a PhD candidate at the University of British
Columbia. His main area of research is in software engineering, with emphasis on web applications, reverse engineering, and software testing. 
He was awarded the four-year fellowship in
2017. He is a student member of the IEEE
Computer Society and served as a part of Virtualization Team in MobileSoft 2020. 
\end{IEEEbiography}

\begin{IEEEbiography}
    [{\includegraphics[width=1in,height=1.25in,clip,keepaspectratio]{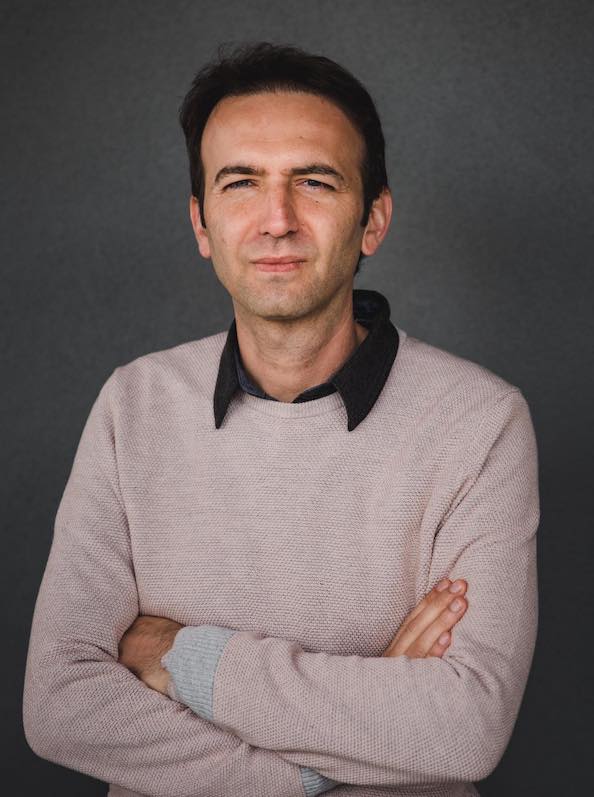}}]{Ali Mesbah}
 is a professor at the
University of British Columbia (UBC) where he
leads the Software Analysis and Testing (SALT)
research lab. His main area of research is in
software engineering and his research interests include software analysis and testing, web
and mobile-based applications, software maintenance and evolution, debugging and fault localization, and automated program repair. He
has published over 60 peer-reviewed papers and
received numerous best paper awards, including
two ACM Distinguished Paper Awards at the International Conference
on Software Engineering (ICSE 2009 and ICSE 2014).  He is the recipient of a Killam Accelerator Research Fellowship (KARF) award (2020), a Killam Faculty Research Prize (2019) at UBC, and was awarded
the NSERC Discovery Accelerator Supplement (DAS) award in 2016. He
is currently an associate editor of the IEEE Transactions on Software
Engineering (TSE) and regularly serves on the program committees of
numerous software engineering conferences such as ICSE, FSE, ASE,
ISSTA, and ICST.
\end{IEEEbiography}

\end{document}